\documentclass[letterpaper,twocolumn,10pt]{article}
\usepackage{usenix2019_v3}

\usepackage{tikz}
\usepackage{amsmath}
\usepackage{graphicx}
\usepackage{xspace}
\usepackage{xcolor}
\usepackage[T1]{fontenc}
\usepackage{subfigure}
\usepackage{verbatim}
\usepackage{url}
\usepackage{enumitem}
\usepackage{listings}
\usepackage{multirow}
\usepackage[ruled,linesnumbered]{algorithm2e}
\usepackage{pifont}
\usepackage[subject={Top1}]{pdfcomment}

\usepackage{filecontents}

\newcommand{\titlesysname}{NetRPC\xspace}
\newcommand{\sysname}{{\small \sf \titlesysname}\xspace}

\newcommand{\para}[1]{\smallskip\noindent\textbf{#1}}

\newcommand{\netlayer}{{\small{\texttt{INCLayer}}}\xspace}
\newcommand{\rpclayer}{{\small{\texttt{RPCLayer}}}\xspace}

\newcommand{\protobuf}{{\small{\texttt{protobuf}}}\xspace}
\newcommand{\netfilter}{{\small{\texttt{NetFilter}}}\xspace}

\newcommand{\pread}{{\small{\texttt{Map.addTo}}}\xspace}
\newcommand{\pwrite}{{\small{\texttt{Map.get}}}\xspace}
\newcommand{\pclear}{{\small{\texttt{Map.clear}}}\xspace}
\newcommand{\pmodify}{{\small{\texttt{Stream.modify}}}\xspace}
\newcommand{\pcntforward}{{\small{\texttt{CntFwd}}}\xspace}
\newcommand{\titlepcntforward}{CntFwd\xspace}
\newcommand{\code}[1]{{\small{\texttt{#1}}}\xspace}

\begin{document}

\date{}

\title{\titlesysname: Enabling In-Network Computation in Remote Procedure Calls}

\author{
{\rm Bohan Zhao}\\
Tsinghua University
\and
{\rm Wenfei Wu}\\
Peking University
\and
{\rm Wei Xu}\\
Tsinghua University
} 
\maketitle

\pagestyle{empty}

\begin{abstract}
People have shown that in-network computation (INC) significantly boosts performance in 
many application scenarios include distributed training, MapReduce, agreement, and network monitoring. 
However, existing INC programming is unfriendly to the normal application developers, demanding tedious network engineering details like flow control, packet organization, chip-specific programming language, and ASIC architecture with many limitations.  
We propose a general INC-enabled RPC system, \sysname. \sysname provides a set of familiar and lightweight interfaces for software developers to describe an INC application using a traditional RPC programming model. 
\sysname also proposes a general-purpose INC implementation together with a set of optimization techniques to guarantee the efficiency of various types of INC applications running on a shared INC data plane. 
We conduct extensive experiments on different types of applications on the real testbed.  Results show that using only about $5\%$ or even fewer human-written lines of code, \sysname can achieve performance similar to the state-of-the-art INC solutions. 
\end{abstract}


\section{Introduction}

The recent programmable switches like Barefoot Tofino \cite{Tofino} and Cisco Silicon One \cite{Silicon} can execute user-specified stateful packet processing at line rate. The evolution has sparked a surge of proposals to offload application functions into the network. The trend is called \emph{in-network computation (INC)}.

INC has been widely applied in various applications including distributed ML training~\cite{sapio2017daiet, graham2016sharp, sapio2019switchML, 265053ATP, viswanathan2019mlfabric}, cache~\cite{jin2017netcache, liu2019distcache}, agreement~\cite{dang2020p4xos, jin2018netchain, yu2020netlock}, and network monitoring~\cite{jepsen2018line, miao2017silkroad, narayana2017monitor, gupta2018sonata}.  The tremendous bandwidth and low latency on switches lead to huge performance gains. For example, ATP~\cite{265053ATP} accelerates distributed training throughput by $38\% \sim 66\%$; P4xos~\cite{dang2020p4xos} reduces the end-to-end delay by more than 90\%; NetCache~\cite{jin2017netcache} improves throughput by 3-10 times compared with a host-only software solution.   
However, developing INC applications involves too much arcane knowledge in networking that is far from application programmers’ (we refer to them as \emph{users} in this paper) expertise and willingness to learn.  

First, the INC program centers on individual packets. Users need to handle network functions such as packet parsing, flow table installation, forwarding, routing, reliable transmission, and congestion control as part of the application.  

Second, users need to learn chip-specific languages like P4 \cite{bosshart2014p4} and NPL \cite{NPL}. Even the high-level programming models like Lyra \cite{gao2020lyra} and C3 \cite{karlos2021C3} still focus on packet processing and require too much network knowledge  (e.g., transmission windows and protocol fields) for software engineers.

Third, users need to understand low-level chip design details and limitations. Familiar data types and operations like floating points are missing, and users have to design approximations~\cite{horrocks2004swrl, 265053ATP, sapio2019switchML} manually. Even harder,
users need to place their program on a pipeline of \emph{stages} with isolated memory and deal with limitations, like once-only memory access per stage and a limited number of tables and rule entries.  

Last but not least, users need to \emph{statically} decide switch memory layout, table/register arrangement, etc., as the switch hardware can only modify them at boot time. Therefore, users need to reset the switch to start/remove an INC application, causing minute-level service interruption.  

As a result, existing projects use INC only as a single application accelerator instead of a shared infrastructure. Even a simple application involves thousands of lines of code on both switches and hosts (Table~\ref{tab:loc} in Section~\ref{sec:evaluation}). The development and operation difficulties prevent wide INC adoption.

In comparison, traditional software uses two abstraction layers to decouple application code from network details:
1) a \emph{Socket layer} providing connection/session management, resource sharing,  reliable communication, and byte stream abstraction; and 2) a \emph{remote procedure call (RPC) layer} providing high-level data types and call interfaces.  
In the popular gRPC framework~\cite{gRPC}, users write a language-independent \emph{interface definition language (IDL)} (e.g., \protobuf~\cite{Protobuf}) specifying types of parameters and return values, and the gRPC compiler generates client and server \emph{stubs} that users can integrate into application code.  The stubs automatically marshal/unmarshal arguments and handle underlying Socket connections. RPCs prove to be a powerful interface to build modern distributed systems.
Unfortunately, neither layer exists in INC, leaving tedious network details to user applications.

We propose \sysname to add both missing layers --- an \emph{\netlayer} and an \emph{\rpclayer} --- to bridge application programming and network packet processing, allowing users to leverage INC features to develop a diverse set of distributed applications using the familiar RPC interfaces.

The \rpclayer provides a high-level RPC interface.  It is built on gRPC with two extensions: \emph{INC-enabled data types (IEDTs)} and a \netfilter. 
IEDTs include basic types like integers and floating points, and collections like arrays and maps.   
Users define RPC services using the same \code{protobuf} language, just replacing vanilla gRPC types with  IEDTs to allow \sysname to recognize and process these data fields. 
In addition, users provide a \netfilter to specify the computation with INC, in terms of five \emph{reliable INC primitives (RIPs)}. RIPs implement high-level operations on IEDTs such as arithmetics, reading/writing a map/array of arbitrary size, and synchronization primitives. RIPs also guarantee reliability, i.e., under various network conditions, RIPs eventually complete as long as the client/server processes survive.

\rpclayer also provides automatic data parallelism for calls with large arguments. \sysname breaks up a call into subtasks, executes these subtasks concurrently, and sends out multiple \emph{concurrent flows}. We offer it as a built-in feature to save programmers from handcrafting concurrent flows or co-flows to fully utilize the 100+ Gbps links in INC switches. 

Analogous to the Socket, the \netlayer handles all flows from the \rpclayer. In addition to the basic guarantees of the Socket-like connection, reliable transport and congestion control, the \netlayer implements the RIPs using a set of protocols involving both the INC switches and the end-hosts.  

We build \sysname as a \emph{general} INC-enabled RPC system.  This is different from existing INC projects that only need to find one workaround for the switch hardware limitations as they target only a single application. The first design tradeoff we need to make is between \emph{generality (i.e., how programmable the network is)} and \emph{simplicity (i.e., how easy it is to program it)}.  
Instead of building yet another general INC language, \sysname chooses to provide only the necessary set of network-independent primitives and the simple \netfilter specification.  
Observing INC projects in the past ten years, we find only a handful successful types (Section~\ref{sec:categories}). We design the primitives so that users can easily develop applications of all these types and enjoy the INC performance boosts.

New challenges for \sysname include, from low level to high level: 1) efficiently managing the switch memory and pipeline stages to support the high-level array and map types; 2) hiding the switch hardware limitations from high-level programs; 3) supporting reliable transmission for different INC scenarios; 4) running multiple INC applications concurrently on a shared data plane; and 5) allowing users to define INC operations for their applications in the familiar gRPC abstraction.  

We have many innovative designs to solve the above challenges.  
1) Using a fallback mechanism, the end-host agents can take over all cases that the INC switches fail to handle; 
2) Using an INC-compatible transport protocol, we can correctly handle packet retransmission and congestion control, maintaining both correctness and throughput;   
3) Adapting a novel memory management scheme, we map from \emph{keys} to unified 32-bit \emph{logical addresses} that further map to switch \emph{physical addresses}, allowing us to optimize the switch memory management much like normal caches;
4) By providing only a limited interface \netfilter, we abstract all obscure hardware limitations into a single high-level limitation (i.e., the primitives \netfilter supports).   

We implement \sysname using a testbed with two Barefoot Tofino \cite{Tofino} switches and eight machines. Using four non-trivial applications (Paxos, network monitoring, distributed training, and MapReduce) as examples, we show that 1) we reduce the \emph{line of code} (LoC) on the end host to about 1/20, using less than two dozen network-related LoC per application; 2) \sysname code is completely the same as vanilla gRPC code; and 3) we can offer the same or even better INC speedup.  

In summary, our contributions include: 

1) As a programming interface, \sysname is the first framework to integrate INC acceleration into the RPC framework, reducing the bar of INC adoption in software.

2) As an INC system, \sysname proposes a set of INC primitives applicable to different INC application types and innovative design elements to efficiently implement them, including reliable transport, memory management, and synchronization, as well as enabling a multi-application INC data plane. 

3) Using four common INC application types on a real testbed, we demonstrate that we can offer the same INC performance boost with far fewer lines of code.

\section{Related Work}
Most existing INC applications make a network-software co-design. Even with the ``network programming languages'', users still have to handle many network engineering details.

\para{Network-software co-design of INC.} 
    People have recently demonstrated many promising INC-accelerated applications, such as NetCache \cite{jin2017netcache} and distCache \cite{liu2019distcache} for caching, P4xos \cite{dang2020p4xos}, NetChain \cite{jin2018netchain} and NetLock \cite{yu2020netlock} for agreement, SwitchML \cite{sapio2019switchML}, SHARP \cite{graham2016sharp}, and ATP \cite{265053ATP} for distributed ML training, and ElasticSketch~\cite{yang2018elastic}, SilkRoad \cite{miao2017silkroad} and Sonata \cite{gupta2018sonata} for network monitoring. These solutions are similarly constructed as the network-software co-design --- user interfaces, customized protocols, switch programs, rule installation, and endpoint agents  ---
    to achieve full-stack optimization and higher switch resource efficiency.

\para{Chip-specific Programming Languages. }
People have proposed several chip-specific programming languages \cite{song2013protocol, sivaraman2016packettrans, bosshart2014p4, NPL} to support data plane customization. Existing programming languages are tightly coupled with corresponding ASICs. For example, Trident-4 \cite{NPL} only supports NPL, while P4 programs can run on Tofino and Silicon One. P4 \cite{bosshart2014p4}, arguably the most popular one for recent INC solutions, follows a \emph{reconfigurable match table} (RMT) architecture. 
P4 programs first define packet headers and corresponding parsers and then process extracted header fields in a pipeline. Programmers must specify the actions on header fields, persistent switch registers at each pipeline stage, and drive actions by match-action tables. Also, users must define a \emph{deparser} to reconstruct the packet for forwarding.
    
\para{High-level network programming abstractions.}
There have been efforts to simplify the INC programming. E.g., Lyra \cite{gao2020lyra} offers a one-big-pipeline abstraction that allows programmers to express their intent with simple statements; NCL \cite{karlos2021C3} imports a window-based abstraction over packets as the basic processing units. $\mu$P4 \cite{soni2020microP4} provides a lightweight logical architecture that abstracts away the structure of the underlying hardware pipelines for better program composition. NetVRM \cite{zhunetvrm} allows developers to virtualize switch memory with a few modifications to existing P4 code.
Chipmunk \cite{gao2019chipmunk} adopts a domain-specific program synthesis technique to generate faster packet-processing code at the cost of longer compilation time. 
However, these high-level abstractions still revolve around networking details, such as (de)packetization, connection maintenance, and protocol stacks.
The semantic gap between the software and network programming model is still a significant obstacle for ordinary software developers. 

\section{Design Overview}
\label{Overview}

We design \sysname to allow software developers to enjoy the performance benefits of INC without tedious network programming. We want \sysname to be general enough to support typical INC application scenarios.

\subsection{INC Application Types}
\label{sec:categories}

\begin{table*}[tb]
	\caption{Four Common INC Application Scenarios and Primitives They Need}
	\label{tab:usecase}
	\small
	\centering
		\begin{tabular}{|c|c|c|c|}
			\hline
			\textbf{Type} & \textbf{Applications and Existing Systems}            & \textbf{IEDT} & \textbf{Primitives}     \\ \hline
			SyncAgtr            & Distributed ML training (ATP~\cite{265053ATP}, SHARP~\cite{graham2016sharp}, SwitchML~\cite{sapio2019switchML})               & Array                   & \pwrite, \pread, \pclear, \pcntforward \\ \hline
			AsyncAgtr           & MapReduce (ASK~\cite{ASK}, NetAccel~\cite{lerner2019netAccel}, Cheetah~\cite{tirmazi2020cheetah})           & Map                     & \pwrite, \pread, \pmodify        \\ \hline
			KeyValue                & Cache (NetCache~\cite{jin2017netcache}, DistCache~\cite{liu2019distcache}), Monitoring (ElasticSketch~\cite{yang2018elastic}) & Map                     & \pwrite, \pread               \\ \hline
			Agreement            & Synchronization (P4xos~\cite{dang2020p4xos}, NetChain~\cite{jin2018netchain}, NetLock~\cite{yu2020netlock})           & Integer                 & \pwrite, \pread, \pclear, \pcntforward \\ \hline
		\end{tabular}%
\end{table*}

INC accelerates applications primarily in two ways: optimizing bandwidth usage (reducing the \emph{number of bytes} to servers) or reducing latency (removing the server from the round trip). People have proposed many INC applications. Table~\ref{sec:categories} summarizes the four types of applications. 

The first two types handle large data sets with optimizing bandwidth as the main goal:  (1) synchronous aggregation (SyncAtgr) for distributed machine learning (ML) training; (2) asynchronous aggregation (AsyncAtgr) for general MapReduce-type applications. The difference between these two types is that SyncAtgr aggregates only a fixed-sized array (e.g., the gradient updates) and works in iterations, i.e., we can proceed only \emph{after} all clients send the updates. In contrast, AsyncAtgr aggregates over an arbitrary number of keys as they come in and allows accessing results at any time.  

The other two types only use small data, with the main goal to optimize latency by avoiding sending packets to the server:  (3) key-value cache (KeyValue) that require frequent queries and responses; and (4) Voting (Agreement) that involve counting votes from different clients until reaching a threshold. Unlike (1) and (2), each request is small, but the challenge is how to achieve a latency smaller than client-to-server RTT by not involving the server at all.

\subsection{Challenges and Solution Overview}
\label{sec:challenge}

\para{Providing a reliable data stream for general INC application types.  }
Different from traditional networks, there are \emph{side effects} when packets go through an INC switch, such as updating a map. Thus, when a packet goes through a switch twice in retransmission, the computation is no longer \emph{idempotent}, violating the computation correctness.
Prior solutions are application-specific, e.g., ATP~\cite{265053ATP} requires explicit server ACKs. It works in SyncAtgr, but not in the other three types because involving the server defeats latency optimization. 
We design an efficient and general retransmission mechanism that maintains the per-flow state on the switch using only a few bits in switch memory. We also design an effective flow and congestion control protocol (Section~\ref{sec:reliable_ds}).

\para{Making ``normal path'' efficient: Supporting memory-efficient arrays and maps on INC switches.  }
Arrays and maps are core data structures in many applications, and INC significantly accelerates operations on them with parallel element processing. E.g.,  training applications use arrays to store the aggregated gradients, and monitoring applications keep the aggregates in a map, one key per metric. In both cases, the switch can add up all values in parallel.
Existing systems either require pre-determined encoding of keys (e.g., knowing all the keys at compile time) or waste precious switch memory and packet header space to store the long keys. We leverage the host agents to generate a \emph{two-level mapping} from keys of arbitrary lengths to a unified 32-bit \emph{logical address space} and then map it to the switch physical memory. We also design a cache management algorithm running on the server agent to improve switch memory utilization efficiency (Section~\ref{sec:memory}).  

\para{Making ``corner cases'' correct:  Hiding switch hardware limitations from the upper-level program.  }
We still need to handle switch hardware limitations.  
Our key idea is to use all \emph{host agents} as a fallback mechanism.  
The host agents emulate all switch operations in software and thus can always provide correct INC results to the \rpclayer regardless of the switch's ability or resource.
\sysname supports two kinds of fallbacks: 1) arithmetic overflows that may happen in floating-point computation and accumulations (Section~\ref{sec:overflow}); and 2) insufficient memory on the switch (Section~\ref{sec:memory}).  

\para{Supporting multi-application data plane.  }
Prior arts support only a single application, and the life span of the switch program does not exceed that of the application. However, the RPC servers are long-running daemons, and server start/stop/restart events are common. It is prohibitively expensive to reset the switch on such events. We solve the problem with three designs: 1) letting all applications share the same set of RIPs; 2) sharing the same set of switch memory blocks among applications by partitioning the key spaces among them; 3) providing three choices of memory eviction behaviors to fit different applications (Section~\ref{sec:memory}). 

\para{Interface \netlayer primitives with \rpclayer without breaking \protobuf abstraction.  }
Users need to tell \sysname \emph{what} to process in INC and \emph{how} to process them. 
We need to add the INC specification to \protobuf language, but we decide \emph{not} to change the language to keep the learning curve low for users. 
Thus, we design the \netfilter as a configuration instead of a program. We only allow users to 
specify a fixed set of RIPs with at most one instance for each kind as a filter to process arguments and return values. The limitation simplifies \sysname design yet still allows implementing all four common types of INC applications (Section~\ref{sec:RPC}).

\begin{figure}[tb]
    \centering\includegraphics[width=1.0\columnwidth]{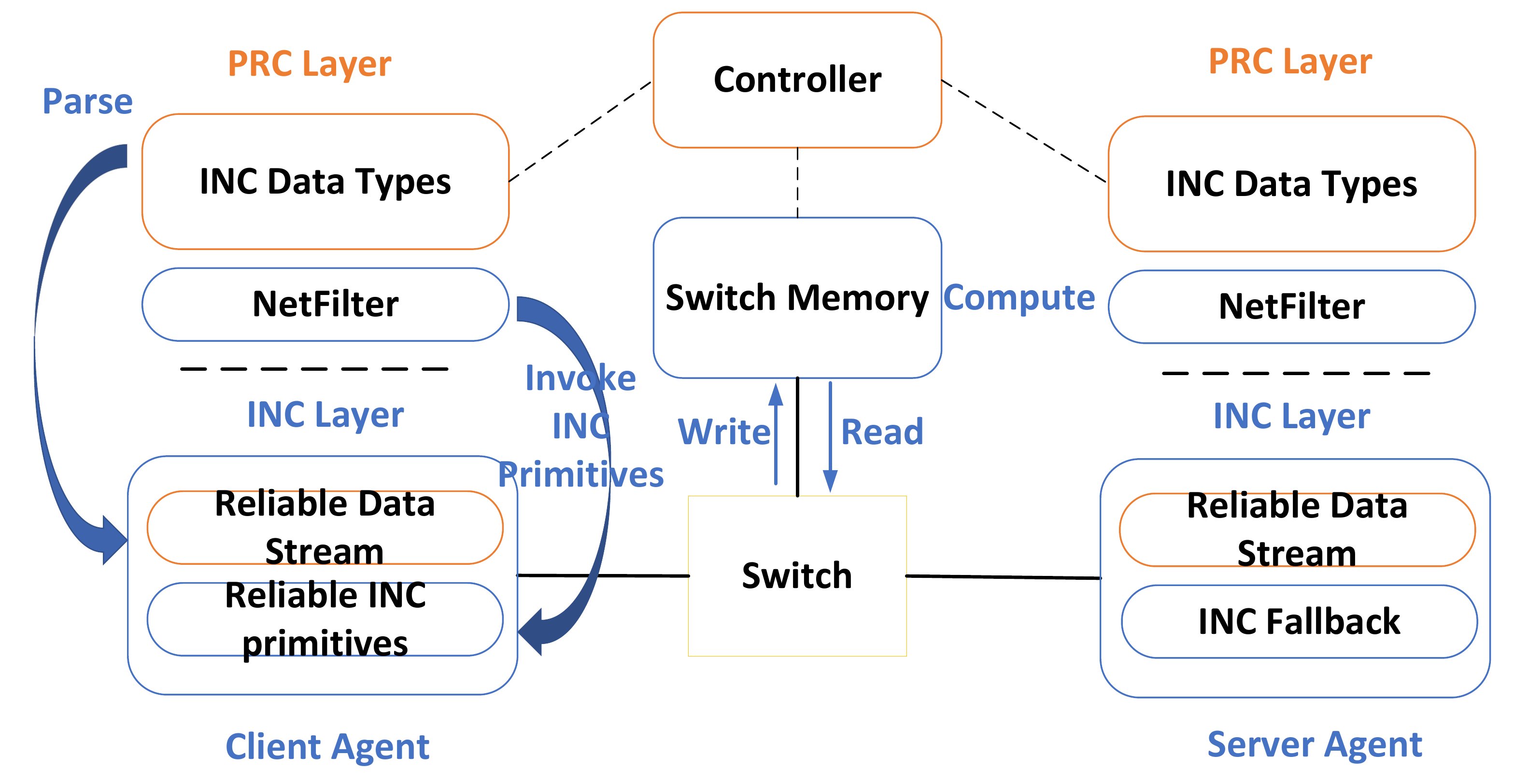}
    \caption{\titlesysname system architecture.} 
    \label{fig:arch}
\end{figure}

The \sysname contains a controller, host agents, and switch programs as in Figure~\ref{fig:arch}. The system-wide \emph{controller} is a dedicated process that handles registration and name lookups at initialization, while at runtime, it manages configurations on both switches and host agents.  
The \emph{host agents} run on each client/server.  
Each host agent maintains a fixed number of connections (configurable) with the switch, even without running tasks. These connections are essential for the reliable communication (Section~\ref{sec:reliable_ds}). 
A single \emph{switch program} starts each INC switch at boot time and executes all primitives. The switch receives configurations from the controller to run applications without resetting the switch program (to avoid interrupting the network). 
If the switch fails to handle a primitive due to resource or functionality limitations, the primitive execution falls back to the server agents.

\section{RPC Layer in \titlesysname}
\label{sec:RPC}
In this section, we first introduce the \sysname programming interface using gradient aggregation in the distributed training application as a concrete example. Then we briefly introduce interface implementation in the \rpclayer. 


\begin{figure}[tb]
\centering
\setlength{\abovecaptionskip}{0.cm}
\begin{minipage}[t]{0.8\columnwidth}
\begin{lstlisting}
import "netrpc.proto"
message NewGrad {
  netrpc.FPArray tensor = 1;
}
message AgtrGrad {
  netrpc.FPArray tensor = 1;
}
service Training {
  rpc Update(NewGrad) returns (AgtrGrad) {} filter "agtr.nf"
}
\end{lstlisting}
\end{minipage}
\caption{Example \protobuf: gradient updates}
\label{code:define}
\end{figure}

\begin{figure}[tb]
\centering
\setlength{\abovecaptionskip}{0.cm}
\begin{minipage}[t]{0.8\columnwidth}
\begin{lstlisting}
{ //agtr.nf
  "AppName": "DT-1",
  "Precision": 8,
  "get": "AgtrGrad.tensor",
  "addTo": "NewGrad.tensor",
  "clear": "copy",
  "modify": "nop",
  "CntFwd": {
    "to": "ALL", 
    "threshold": 2, 
    "key": "ClientID",
  },
}
\end{lstlisting}
\end{minipage}
\caption{Example \netfilter: gradient updates}
\label{code:conf}
\end{figure}

\begin{figure}[tb]
	\centering
	\setlength{\abovecaptionskip}{0.cm}
\hfill
\begin{minipage}{0.92\columnwidth}
	\begin{lstlisting}
shared_ptr<Channel> channel = CreateCustomChannel(server_ip, InsecureChannelCredentials());
unique_ptr<Stub> stub_(NewStub(channel));
void PushPull(double* data, int length) {
  NewGrad request;
  AgtrGrad reply;
  ClientContext context;
  request.mutable_tensor()->mutable_data()
        ->Add(data, data+length);
  Status status = stub_
        ->Update(&context, request, &reply);
  memcpy(data, reply.tensor().data(), 
        length * sizeof(double))
  train(data);
}
	\end{lstlisting}
\end{minipage}
	\caption{Client program to use the RPC}
	\label{code:example}
\end{figure}

\para{\protobuf definition.  }
Like in vanilla gRPC, users first provide a \protobuf definition that compiles into the client and server stubs. 
Figure~\ref{code:define} shows an example \protobuf file.  The \texttt{message}s are user-defined types, and \code{service} is the RPC definition using \texttt{message}s as arguments and return values.  The only modification to vanilla \protobuf is the \code{filter} clause allowing users to provide the \netfilter file name (see below). 

\para{\sysname data types.  }  Users declare all variables that they want to process in INC using \emph{INC-enabled data types (IEDTs)} defined by \sysname. E.g., line 3 and 5 in Figure~\ref{code:define} defines variables (both \code{tensor}) as a \code{netrpc.FPArray} (floating point array) IEDT.  Optionally, user can add normal gRPC data fields to the same messages, and \sysname simply passes them to the server without processing.  

Collections (\code{Array} and \code{Map}) are core data types in \sysname.  The item value can be integers or floating points, and keys can be integers or strings.  
\sysname enables 1) automatically applying the user-defined \netfilter on \emph{every} value in these collections and 2) accessing the \emph{global INC map} using keys.

\para{Life of a \sysname call. }
In \sysname, when a client initiates a call, the \emph{client stub} marshals the arguments and sends them through one of two channels: messages with IEDT through the INC channel established by the per-host \emph{client agent} and normal messages through the original gRPC Socket. In this paper, we only focus on the \emph{data streams} in the INC channel. The underlying \netlayer processes the data stream and optionally interacts with the \emph{INC map}. The INC map is a \sysname abstraction of unlimited global memory addressable using keys or array indices. INC map is implemented on both switches and host agents (in Section~\ref{sec:memory}).
The return path is similar: the \emph{server stub} marshals the return value and sends it through either the INC channel or the normal Socket.

\begin{table}[tb]
	\caption{\sysname Primitive Semantics}
	\label{tab:primitive}
	\resizebox{\columnwidth}{!}{%
		\begin{tabular}{|c|c|c|}
			\hline
			Primitive     & Args       & Semantics                                                                                                                           \\ \hline
			\pread     & stream     & map{[}stream.key{]} += stream.value                                                                                                 \\ \hline
			\pwrite       & stream     & stream.value = map{[}stream.key{]}                                                                                                  \\ \hline
			\pclear     & empty      & map{[}stream.key{]} = 0                                                                                                             \\ \hline
			\pmodify & op,para  & stream.value = op(stream.value, para)                                                                                               \\ \hline
			\pcntforward       & key,th,tgt & \begin{tabular}[c]{@{}c@{}}cnt{[}key{]}++; if cnt{[}key{]} == th \\ then forward(tgt) else drop\end{tabular} \\ \hline
		\end{tabular}%
	}
\end{table}

\begin{figure}[tb]
    \centering\includegraphics[width=1.0\columnwidth]{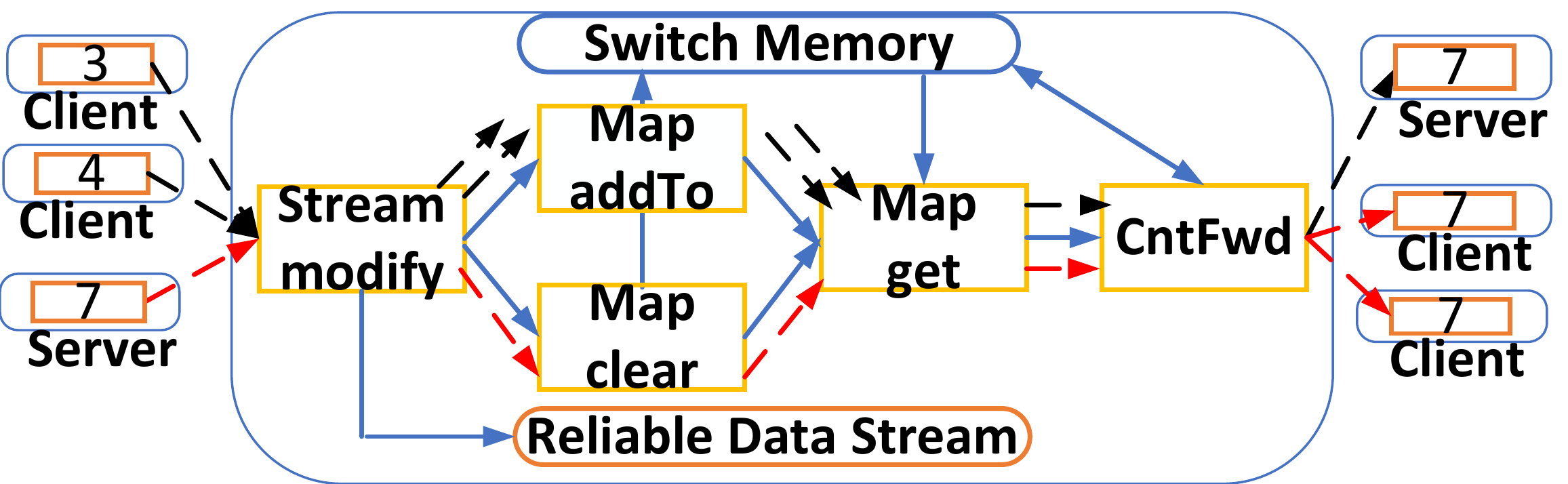}
    \caption{RIP pipeline in switches.} 
    \label{fig:workflow}
\end{figure}

\para{The \netfilter and reliable INC primitives (RIPs).  }
In addition, users need to specify their INC operations.  
Here, we have a choice in terms of what kind of operations \sysname should provide. We want to find the sweet spot in the trade-off between generality and simplicity. We also want to provide a reconfigurable switch program to serve new applications. Therefore, we pick five primitives that we can compose together in a similar layout to implement existing types of INC operations (Section~\ref{sec:categories}). Figure~\ref{fig:workflow} displays this layout and its implementation on the switch. The users only need to provide configurations for these five primitives in their \netfilter file  (Figure~\ref{code:conf}) to specify their INC operation of interest.

The \netfilter is a JSON configuration file. It contains a \code{AppName} that uniquely identifies an application, a \code{Precision} field that specifies the floating-point precision (number of digits after the decimal point). Lower precision allows INC to process more data without falling back to the host.  

The more interesting part in \netfilter is the next five fields that allow users to provide arguments to RIPs, including three map-access primitives, \pread, \pwrite, and \pclear, one data stream manipulation primitive, \pmodify, and one synchronization primitive, \pcntforward. Table~\ref{tab:primitive} summarizes the parameters and semantics of these primitives. 

\pread \emph{accumulates} data items from the stream to the map according to their keys/indices, and \pwrite reads out the values of a specific key from the map. In Figure~\ref{code:conf}, we add the values of the \code{NewGrad.tensor} array to the INC map to aggregate the gradient, and on the return path, we read out the results from the INC map into the \code{AgtrGrad.tensor} array.  

\pclear defines how to clear a value from the INC map. In the example, \code{copy} means backing up the aggregates to the server before clearing it out to handle packet losses.  
We introduce other possible options in Section~\ref{sec:memory}.  

\pmodify performs arithmetics on the stream. It only modifies the stream without accessing the INC map. In Figure~\ref{code:conf}, we set it to \code{nop}, as we do not modify streams.  Table~\ref{tab:op} in Appendix~\ref{sec:op} lists all operations we support for \pmodify. 

The \pcntforward is the most interesting primitive. It accumulates values on one or more keys (specified with \pcntforward\code{.key}) in the INC map until the accumulator reaches the specified threshold (\pcntforward\code{.threshold}). Then it forwards out the message to the destination(s) specified at \pcntforward\code{.to}.  
The \pcntforward primitive is essential to control both \emph{how many} packets to forward to the clients/servers and \emph{when} to forward them, and thus essential for SyncAgtr and Agreement applications.  
In this example, we set \code{key} to a single \code{ClientID}, meaning that we only need one counter for the number of unique clients who have sent gradient updates. In this case, only when exactly two unique clients have sent a stream, will the network aggregate the items and send back \code{AgtrGrad} to \code{ALL} clients.  

There are other use cases for \pcntforward. Setting the \pcntforward\code{.threshold} to one makes the \pcntforward behave as the \code{test\&set} primitive in many instructions sets, useful to implement distributed mutual exclusion.  
Also, by providing a collection in the data stream, we can use a map of counters to track multiple votes in concurrent ballots, a widely-used functionality in distributed agreement protocols.  \pcntforward allows the switch to notify the clients only when enough votes arrive. Appendix~\ref{appendix:example} provides more examples of \pcntforward primitive. 

Table~\ref{tab:usecase}  summarizes the primitives used in each INC application type. 
Figure~\ref{fig:workflow} illustrates a RIP pipeline running the example code in Figure~\ref{code:conf}. A SyncAgtr application pushes data into the network by its clients (black arrows) for on-switch aggregation and on-server backup. The server sends back the computation results (red arrows) to clients and clears the switch memory. The same switch program (and the RIP pipeline) completes all INC functions in the round trip without reconfiguring the switch. 


\para{Using RPC calls to build the application.}
With the \protobuf and \netfilter definition, the remaining process is exactly the same as normal gRPC. The \protobuf compiler generates client and server stubs, and users include stubs in their application.  
Figure~\ref{code:example} provides the client code using the RPC service defined in Figure~\ref{code:define}. Note that the code is completely identical to vanilla gRPC, hiding INC details from the users.

\para{Automatic data parallelism on RPCs with large arguments.  }
\label{sec:parallel}
There could be multiple concurrent \sysname applications/channels and procedure calls in the runtime. 
The stub submits calls as tasks to the host agents. A task contains the application data (i.e., arguments or returned values, encoded as protobuf messages), the metadata (e.g., network program configurations), and the routing information.

The host agents maintain a thread pool of worker threads to process tasks. \sysname automatically partitions the task into subtasks and dispatches them onto multiple worker threads for load balance. 
The worker threads serve the subtasks in their queue on a First-Come-First-Serve (FCFS) basis. The worker threads serialize the subtask's data into a sequence of packets (Appendix~\ref{sec:header}) and send them over the user-level network stack we implement using DPDK~\cite{DPDK}.

\para{Limitation of RIP abstraction.}
RIP abstraction targets simplifying programming in general INC scenarios where different INC applications regularly start and stop to share the infrastructure. It lacks logical semantics like looping and branching and thus can not implement complex algorithms (e.g., DHS~\cite{zhao2021dhs}) or data structure (e.g., NetChain~\cite{jin2018netchain} queue). 
Adding more RIPs will extend the functionality but reduce the source available for each RIP. We leave further extensions and a customizable set of RIPs as future work.

\section{INC Layer in \titlesysname}

\netlayer provides a reliable layer to support RIPs. There are two design objectives: 
1) efficiently utilize INC switch resources to support a multi-application data plane with full INC performance boost; 
2) provide an end-host software-based fallback mechanism to support reliable byte streams and INC primitives for the \rpclayer. 
As a result, \rpclayer can safely assume that the data stream is delivered reliably, and the \netfilter is fully executed in various network conditions.  




In this section, we first introduce how we can build a general reliable data stream abstraction. Then we introduce the essential RIPs, including map access, arithmetics, and \pcntforward. Finally, we briefly introduce the switch implementation.

\subsection{Reliable Data Stream }
\label{sec:reliable_ds}

\para{Encoding IEDTs into sequences of packets.  }
Client agents receive data streams containing multiple key-value pairs from the \rpclayer.
Then the client agent encodes these pairs into separate packet headers using a user-level networking stack written in DPDK~\cite{DPDK} and sends them out.  
Each packet contains a fixed number of key-value pairs (32 in the current setting), a sequence number, a \emph{global application ID} (GAID), as well as other state information we will introduce in this section. Figure~\ref{fig:header} in Appendix~\ref{sec:header} illustrates the packet header structure.  The packet can optionally contain the normal payload with non-INC types for the application. \sysname only processes the key-values encoded in the header.

\para{Idempotent packet retransmission.  }
In case of packet loss, traditional transport simply retransmits the lost packet. However, INC complicates the retransmission because the retransmitted packet can cause \emph{side effects} on the switch, such as incrementing a map value again. In other words, naive retransmission is not \emph{idempotent} and may lead to incorrect INC results.  
Switches need extra information to detect which packets are retransmitted.
Traditional networking doctrine tells us that we shall avoid keeping states on switches. Thus,
some INC designs choose to keep extra states on the sever~\cite{265053ATP} and let the server to ACK each packet. The ACK informs the switches about processed packets. This design requires the server to ACK every packet. 
It works on applications like gradient aggregation, where the INC is primarily used for reducing \emph{server bandwidth} by only forwarding the results to the server to ACK while completing aggregation on the switch only. 
However, this design does not fit applications like KeyValue and Agreement, as server ACKs defeat the purpose of latency reduction via sub-RTT switch response.

To design a general protocol, we observe that 1) we send/receive all INC flows using host-agents under \sysname control; 2) the INC switches have a relatively large memory, and the retransmission states are almost negligible compared to the INC map. Thus, we can safely keep the per-flow state on the switch as long as we can limit the number of agent flows.

We further design our protocol to minimize per-flow switch memory usage, allowing a switch to only keep a \emph{bit array} of size $w_{max}$ per flow, where $w_{max}$ is the max sending window size.  The switch initializes all bits to 1. Every packet contains a sequence number (\code{seq}), and a \code{flip} bit that is set to $(seq / w_{max})\%2$.  
On receiving the packet, the switch checks the $(seq \% w_{max})$-th bit in the bit array. If the bit is the same as the \code{flip} in the packet header, the switch considers it as a retransmitted packet and thus skips updating the INC map. Otherwise, the packet is new, and the switch sets its corresponding bit to the \code{flip} and processes the packet normally.

We show that this simple protocol guarantees idempotent execution, i.e., 1) a packet's first appearance flips the bit, and 2) a packet's later appearance (retransmission) equals the bit. We prove it by \emph{induction}. 
For sending window $0$, all \code{flip} bits in packets are $0$s, and the switch bitmap is all $1$s. Since each packet only sets the bitmap to 0 once, at the end of window $0$, all bits become 0.  
Then assuming the two properties hold for window $t-1$, we show that they still hold for window $t$.  
Recall that the client agent sends out the $i$-th packet in window $t$ (denoted as $P$) only after the $i$-th packet in window $t-1$ (denoted as $P'$) is ACKed.  
Therefore, when $P$ first appears, $P'$ should already set the bit as $P'$'s \code{flip}. 
As $P$ and $P'$ are opposite in \code{flip}, $P$'s first appearance flips the bit. $P$'s later appearance would not flip the bit, and the window controls packets out of the window not to appear (and falsely flip the bit) between $P$'s appearances. Thus, the two properties hold in window $t$. By induction, it is correct for all sending windows.



We can use $N\times w_{max}$ bits of switch memory to support $N$ concurrent reliable flows on each agent. We experimentally set $w_{max}=256$ and find it sufficient to achieve a per-flow bandwidth of 20+ Gpbs.  

\para{Flow control and congestion control.  }
\label{sec:cc}
Note that $w_{max}$ is a fixed value.  We still need to deal with the flow control and congestion control due to resource contentions on either the end hosts or switches. We use the same mechanism to handle both flow and congestion control by automatically setting a congestion window $cw \leq w_{max}$.  

Traditional congestion control, like the one in TCP, relies on round-trip-time (RTT) and duplicate ACKs to adjust $cw$. 
However, in INC primitives like \pcntforward, these signals may not reflect the real network congestion as the receiver needs to wait for the slowest sender before ACKing. 

Thus, \sysname adopts an ECN (i.e., explicit congestion notification)-based congestion control mechanism. The switches set the ECN when the ingress port length exceeds the threshold. Meanwhile, it writes the ECN information to the INC map under a special key. Thus, all retransmission packets carry ECN until cleared like other map values. This prevents ECN signal loss due to packet loss.
Otherwise, the client agents adjust the $cw$ using the same \emph{additive increase multiplicative decrease (AIMD)} policy as priot arts~\cite{265053ATP}. 
Experiments show that this design allows multiple flows to achieve both high goodput and fairness in bandwidth sharing. 

\para{Other transport protocols.}
Several recent transport protocols have affected the design of \netlayer. 
MTP~\cite{stephens2021mtp} proposes a message-based protocol to customize congestion control, load balancing and resource isolation for INC.
However, it requires maintaining per-pathlet states on both packet headers and end hosts, importing extra overhead to hurt the system's performance.  
DCTCP~\cite{alizadeh2010dctcp} imports a more fine-grained congestion window adjustment based on the ECN proportion. 
This approach is inapplicable to INC scenarios because we have to count the maximum number of ECNs in a single (i.e., the most congested) path instead of the total ECN proportion due to incast. This consumes more resources on the switch and will reduce the stream goodput, so we utilize AIMD to simplify implementation and will extend the protocol in future work (Section~\ref{sec:conclusion}).

\subsection{Reliable INC Primitive Designs}

\subsubsection{Computation and arithmetic overflows }
\label{sec:overflow}
\para{Floating point arithmetic by quantization.  }
INC switches only have limited 32-bit arithmetic functionality, yet INC applications like training require floating-point (FP) arithmetic. The standard practice uses quantization to fit the FP numbers into 32-bit integers (aka, fixed-point numbers).  \sysname quantizes an FP value in the client agent by multiplying it with a scaling factor (the \code{precision} field in \netfilter) and maps the value back to FP before handing it to the \rpclayer. 

\para{Handling overflows.  }
While people have shown that the precision loss might not cause problems in many applications,  32-bit fixed-point numbers do not offer enough \emph{representable range} in many cases, and thus overflow is unavoidable. Even without FP numbers, just using the \pread to accumulate values may also lead to overflow. Thus, we need a way to handle occasional overflows. 

When the switch detects an overflow during computation, it sets the overflowed value to \code{MAX\_INT} or \code{MIN\_INT} and forwards the packet normally. 
When a host agent receives a packet with \code{MAX\_INT}/\code{MIN\_INT} value, it suspects there is an overflow~\footnote{Strictly speaking, there is one possibility of a false positive where the result is exactly \code{MAX\_INT}/\code{MIN\_INT}. The false positive only slightly affects performance leading to an extra retry, but not correctness.} and gives up the result. 
Then the client agents mark and resend these overflow packets, causing the switch to skip the processing and directly forward them to the server agent. 
The server agent computes the correct result using 64-bit integers or FP numbers in software.   

\para{Fallback on network fabrics without INC support.} 
A similar fallback mechanism works when there are no programmable switches or data-plane resources reach capacity. If the controller fails to assign the INC application to any switch, the server agent will execute RIPs in software using the same switch failure handling mechanism. Therefore, the application is guaranteed to derive the correct results with the transparent fallback, only losing performance benefits from INC.

\subsubsection{Memory: INC map-access primitives}
\label{sec:memory}

\para{Memory address spaces in \netlayer.} The \rpclayer supports maps with arbitrary keys, while the \netlayer only provides a 32-bit \emph{logical} address space per application. The client agent hashes keys with different types and lengths into the 32-bit address space. We handle all collisions by putting the colliding keys into the payload to bypass the switch INC and let the server agent to process them. We choose not to use a larger logical space as we find it sufficient to support multiple applications with acceptable collisions. A short address saves bits in packets, increasing the effective bandwidth. 

\netlayer maps the 32-bit address space onto the \emph{physical} address space on switches. Each physical address corresponds to a \emph{register} on a switch. Switches may have different numbers of registers. E.g., the switch we use has about 160K registers available per pipeline stage, and we use eight stages to support map-related primitives.  

It is not trivial to map the logical address to a physical switch address. The above hashing approach does not work here because switch registers are a valuable resource we want to make full use of, but when the utilization is high, the collision rate increases fast, causing many fallbacks to servers.  
In fact, we need to pack the physical memory tightly.  
Also, we want to avoid keeping the logical-physical address mapping on switches; otherwise, it wastes switch memory. 

In some applications, such as distributed training (as in ATP~\cite{265053ATP}) in SyncAgtr, it is simple as every client has the same set of keys. Each of them only needs to sort the keys and give each key a sequence number. However, it does not work in general cases, such as AsyncAtgr, where each client might have a different set of keys.  

\para{Multiple clients of a single application.  }
We solve the problem by letting the server agent, shared by multiple clients, decide and maintain the mapping for all its clients. The first time a client uses a new logical address, it sends packets to the server without INC. If there is switch memory available, the server agent will piggyback a mapping for this address on the returning ACK. Then the client can send subsequent packets with the physical address set in the packet for the switch. If the switch memory is full, the server will not return the mapping, and thus clients keep sending subsequent packets to the server without INC. With the method, we ensure all clients calling the same server use a consistent mapping.  

\para{Handling multiple applications.  }
According to the applications' requests, the controller reserves switch memory at application registration time. When an application gets no switch memory, they fall back to using server agents. We use a simple FCFS policy for the static allocation among different applications and leave advanced memory scheduling as future work.  
Note that although the controller reserves memory at registration time, the actual allocation only happens when the clients plan to send out data streams. Thus we can avoid holding memory unnecessarily.

\para{Cache replacement policies.  } 
The switch memory serves as a cache for certain keys, and we need a replacement policy at the server agent. We take an approximation to the \emph{least-recently-used} (LRU) policy. Each client agent counts the uses of each logical address within a \emph{cache update window}, and at the end of the window, they send the counter to the server, allowing the server to compute the most-used keys in the last window. Then in the next period, the server evicts less used values. We also evaluate other popular cache replacement policies in Section~\ref{sec:evaluation}, and we show that this periodic counting-based LRU policy works well.

\para{Optimization for synchronous aggregation.  }
In addition to the general logical-physical mapping, we realize that the SyncAgtr (i.e., distributed training) applications like SwitchML~\cite{sapio2019switchML} only require access to large continuous arrays. It is more memory-efficient to be able to allocate such arrays in a few \emph{circular buffers} instead of many individual addresses.  \sysname supports such buffers of a fixed size of 256 keys.

\para{Preventing switch memory leaks on host failures.  } 
Unlike existing INC designs that serve only a single application, \sysname is a shared infrastructure supporting many applications.   Thus, we need to take care of potential switch memory leaks resulting from the crashing of user programs or host machines before they can explicitly release the memory. We address this issue with a \emph{two-level timeout} mechanism.  

\sysname processes a packet with an \emph{admission rule} that checks the GAID. We keep a timestamp of the last time the rule runs for each GAID.  
The controller periodically polls the switch for these timestamps. If it finds a stale timestamp, it triggers the \emph{first-level timeout} by notifying the server agent to retrieve the application's INC map. After a longer period, the server agent triggers the \emph{second-level timeout}, sending the saved data items to the user-defined stub or deleting them if the stub no longer exists.  
As switch memory is small and precious, we want to reclaim it quickly with a small \emph{first-level timeout}. However, the small timeout unavoidably introduces false positives, hurting the correctness of programs with low communication frequency, such as monitoring infrequent events. In fact, these applications will benefit little from INC anyways, and the timeout mechanism allows them to run just like normal applications. Servers have much larger memory and thus can keep user maps longer, providing the correctness of such programs similar to software.

\para{The \pclear primitive.}
The switch memory only supports \pread instead of directly overwriting the value. Thus, to start a new accumulation (e.g., a new iteration of training, restarting a vote, etc. ), the user program needs to execute three steps: 1) \pwrite the accumulator value to the hosts, and 2) \pclear the memory and 3) start to \pread new values.  
However, there is a risk that the packets get dropped \emph{en-routing} to the host. In this case, the memory is already cleared, so the value is permanently lost.  

\sysname provides different methods to prevent this loss, as there is a latency-throughput tradeoff. We decide to allow users to choose from three clear policies in \netfilter.  

1) [Copy]: The client-call stream first carries the map's value to the server, and then the return stream from the server will \pwrite and \pclear the values. Thus we guarantee the server has a backup in case the return packet is lost. This policy requires no extra switch memory at the cost of forwarding more data to the server and thus higher latency.

2) [Shadow]: The switches double memory allocation. The data stream uses two memory segments alternatively: \pwrite from one and \pclear the other. This approach reduces latency at the cost of doubling memory usage and thus is only suitable for latency-sensitive applications with few data items. 

3) [Lazy]: The \pclear primitive only lets the host agents to save the current value and let the switch to keep accumulating without clearing. The host agent subtracts the saved value to compute the accumulated value since the last clear. When the accumulator eventually overflows, we fall back to the server agent using the same overflow logic and clear the switch memory. If the application (e.g., voting) has a slow-increasing counter, lazy policy involves little overhead. 

The multiple \code{clear} policies allow users to better customize their INC applications according to their SLA requirements and workload features. We compare the performance of the three policies in Section~\ref{sec:evaluation}. 

\para{Implementation on the switch.  }
We allow 32 key-value pairs per packet. We use four register groups per stage and 8 out of the 12 stages on the switch to implement the INC map access. This design fits the switch hardware limitation: a packet can only access each group of registers in the switch once per trip. For the same reason, we arrange \pwrite/\pread and \pclear to execute in the opposite direction of a packet round trip. These primitives are organized in a flow chart on the switch pipeline (Figure~\ref{fig:logic} in Appendix~\ref{sec:switch-pipeline}).
Appendix~\ref{appendix:example} displays a number of example settings of \netfilter in different application types.

\subsubsection{Forwarding: the \titlepcntforward primitive }
The \pcntforward primitive requires two extra pieces of logic in the switch. First, the switch needs to recognize the packet is a \pcntforward packet, and then the packet goes through the normal map-access pipeline to increase and read the values in the accumulator. 
We implement different computation logic for the accumulator (\code{test\&set} or accumulate) by applying different match-action tables according to the \pcntforward\code{.threshold}. 
Finally the packet enters the last stage on the switch that decides whether to \code{drop}, \code{send}, or \code{multicast} the packet.  
\section{Evaluation}
\label{sec:evaluation}

In this section, we show that \sysname achieves the following desirable properties:
1) \sysname supports four kinds of INC applications; 2) \sysname significantly reduces the amount of application code; 3) \sysname achieves the same performance as handcrafted INC applications; 4) \sysname handles situations like packet loss, congestion, etc.  
In addition, we evaluate the effects of policy settings (clear and caching).  

\subsection{Experiment Settings}

\para{\titlesysname implementation.  }
We implement \sysname switch logic on a 12-stage programmable switch. The \sysname switch pipeline contains 32 read-write memory segments corresponding to the 32 key-value pairs in the \sysname packet. 
Each memory segment contains 40k 32-bit units to restore INC states or the INC map. Depending on the service configuration, we vary packet lengths from 192 to 320 bytes. 

\sysname includes four modules: $\sim 4K$ lines of P4 code for the switch logic, $\sim 2K$ lines of Python code for the remote controller, $\sim 2K$ lines of C++ code as the plugin of gRPC++~\cite{gRPC}, and $\sim 3K$ lines of C++ code for the \sysname end-host agents using DPDK. We also implement four types of INC applications with only $200 \sim 500$ lines of code each.

\para{Testbed.} We run \sysname on a testbed of 8 GPU machines and two programmable switches. The devices form a dumbbell topology: two connected switches, each with four machines. In the experiment, we use ``$X$-to-$Y$'' to denote a topology with $X$ clients and $Y$ servers.
The switch contains a Barefoot Tofino chip and provides 32 $\times$      100 Gbps ports. 
Each machine has a Mellanox ConnectX-5 dual-port 100 Gbps NIC.
Each machine is equipped with two NVIDIA GeForce RTX 2080Ti GPUs, 56 CPU cores at 2.20GHz, and 192GB RAM. The machines install NVIDIA driver 430.34, CUDA 10.0, Mellanox driver OFED 4.7-1.0.0.1, and Ubuntu 18.04.

\para{Workloads and baselines.  } Table~\ref{tab:workload} shows the workloads and baselines we use. We run various typical models (VGG, ResNet, AlexNet) for SyncAgtr. We also implement each application's pure software version as baselines using DPDK. 

\begin{table}[tb]
\caption{Workload and Baseline in Experiments}
\label{tab:workload}
\footnotesize
\begin{tabular}{|c|c|c|c|}
\hline
App Type  & App                                                             & INC Baselines  & Dataset                                                                    \\ \hline
SyncAgtr  & \begin{tabular}[c]{@{}c@{}}Distributed \\ Training\end{tabular} & \begin{tabular}[c]{@{}c@{}}ATP~\cite{265053ATP} \\ SwitchML~\cite{sapio2019switchML}\end{tabular}  & ImageNet~\cite{ImageNet}                                                                   \\ \hline
AsyncAgtr & WordCount                                                       & ASK~\cite{ASK}           & Yelp\cite{Yelp}                                                                       \\ \hline
KeyValue  & \begin{tabular}[c]{@{}c@{}}Network \\ Monitoring\end{tabular}   & ElasticSketch~\cite{yang2018elastic} & \begin{tabular}[c]{@{}c@{}}CAIDA Anonymized \\ Internet Trace\cite{CAIDA}\end{tabular} \\ \hline
Agreement & Paxos                                                           & P4xos~\cite{dang2020p4xos}         & Synthatic workload                                                           \\ \hline
\end{tabular}%
\end{table}

\subsection{Reducing User Code Complexity}
\begin{table}[tb]
\setlength{\abovecaptionskip}{0.cm}
\caption{LoC Comparisons: \sysname vs. Prior INC Arts}
\label{tab:loc}
\centering
\small
\begin{tabular}{|c|cc|cc|}
\hline
\multirow{2}{*}{} & \multicolumn{2}{c|}{\sysname}              & \multicolumn{2}{c|}{Prior INC Arts} \\ \cline{2-5} 
                  & \multicolumn{1}{c|}{Endhost} & Switch & \multicolumn{1}{c|}{Endhost}  & Switch  \\ \hline
SyncAggr         & \multicolumn{1}{c|}{173}     & 13      & \multicolumn{1}{c|}{3394}     & 5329     \\ \hline
AsyncAggr        & \multicolumn{1}{c|}{166}     & 26      & \multicolumn{1}{c|}{3278}     & 4258     \\ \hline
KeyValue           & \multicolumn{1}{c|}{162}     & 26      & \multicolumn{1}{c|}{898}        & 2360     \\ \hline
Agreement             & \multicolumn{1}{c|}{1453}    & 26      & \multicolumn{1}{c|}{5441}     & 931      \\ \hline
\end{tabular}%
\end{table}

We compare the user-written lines of code (LoC) of \sysname applications with existing INC arts. Table~\ref{tab:loc} shows that \sysname reduces the overall human-written code by over 97\% in all four application types.
To enable INC in an RPC, the application developers only need to configure the \code{NetFilter} to enable/disable RIPs on the switch without writing any switch code.  \netfilter results in a huge LoC reduction (12-21 LoCs in \sysname v.s. 931-5329 in prior arts). On the host, \sysname also reduces the LoC of host programs by 95\%, 95\%, 73\%, and 82\% for the four applications compared with existing INC applications, as \sysname users only write code to process data-stream as call arguments, avoiding the tedious network functions like (de)packetization, reliability, etc. 


\begin{figure*}[htb]
\begin{minipage}{0.37\textwidth}
\includegraphics[width=1.0\columnwidth]{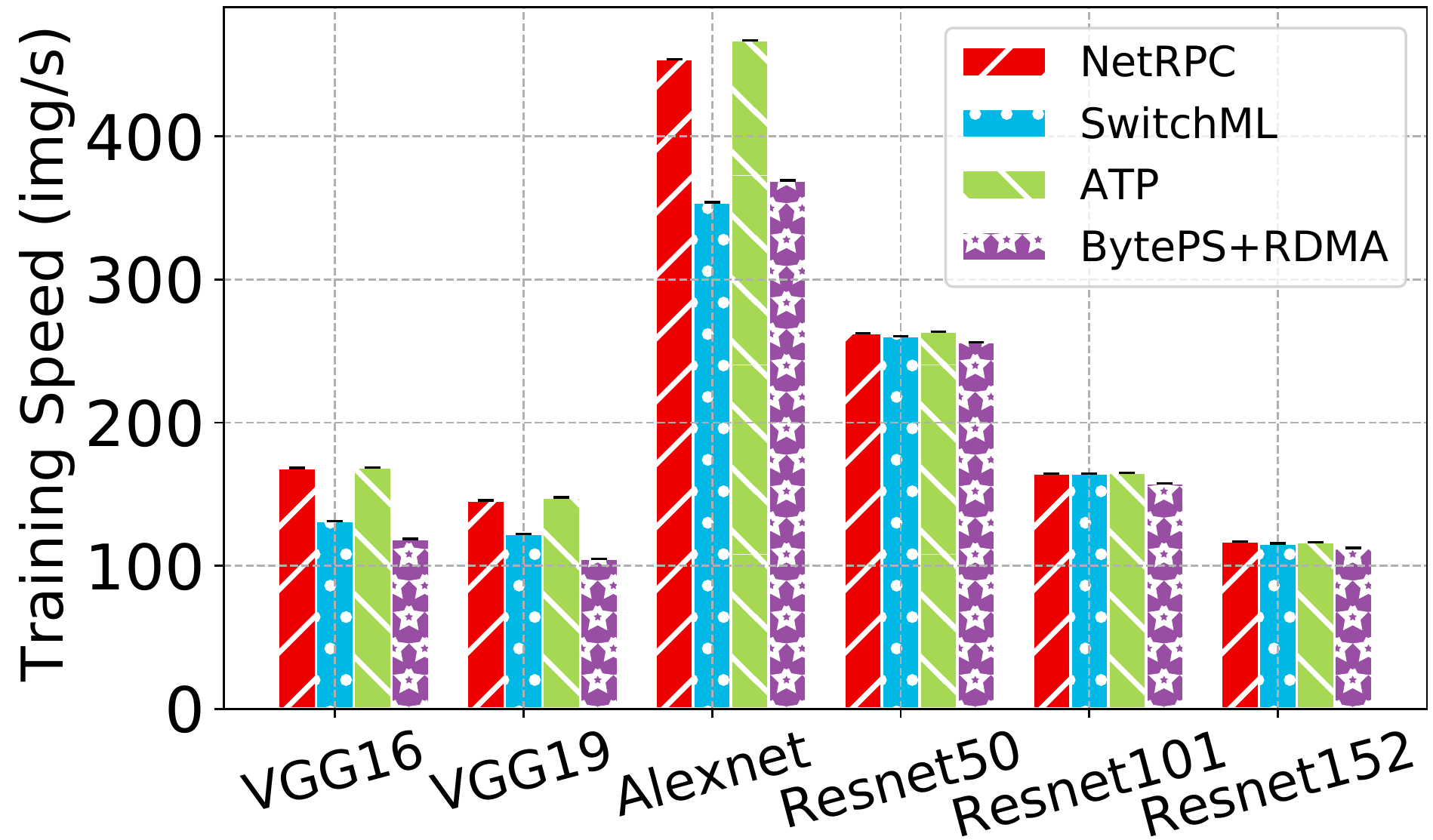}
\centering
\caption{Deep Learning Training Speed}
\label{fig:byteps}
\end{minipage}
\hfill
\vspace{0pt}
\begin{minipage}{0.6\textwidth}
    \subfigure[\small 99th-Percentile Latency]{
    \normalsize
    \includegraphics[width=0.49\linewidth]{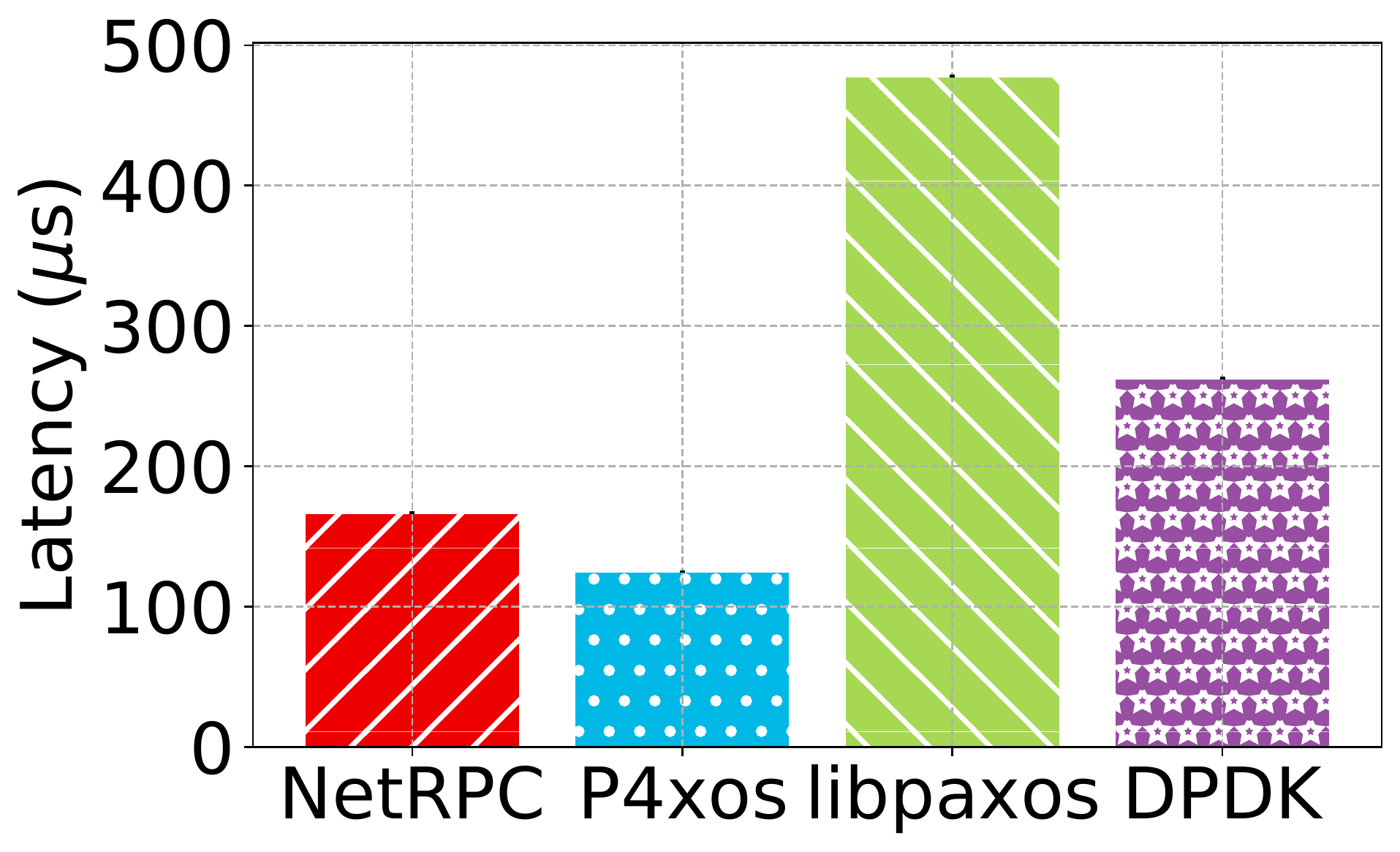}
    }%
    \subfigure[\small Throughput]{
    \normalsize
    \includegraphics[width=0.49\linewidth]{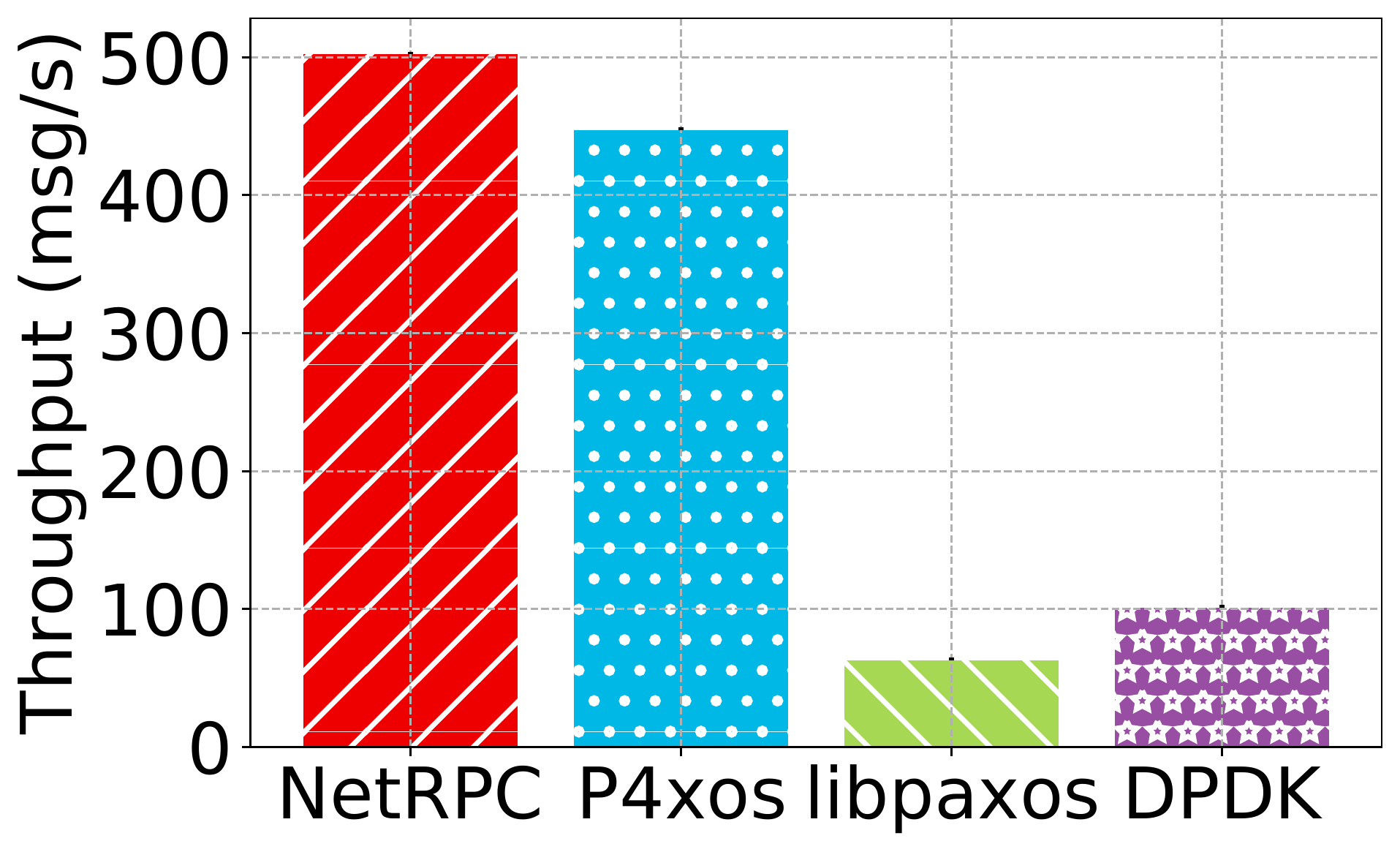}
    }%
    \centering
    \caption{End-to-end Performance of Paxos Systems.}
    \label{fig:paxos}
\end{minipage}
\end{figure*}

\subsection{End-to-end Application Performance}
\label{sec:macrobenchmark}

\para{Distributed ML training.} 
We set up eight worker machines for this evaluation. We use two existing INC frameworks, SwitchML~\cite{sapio2019switchML} and ATP~\cite{265053ATP}, and a pure software solution, BytePS, as baselines. We implement the \sysname version on BytePS with only 500 LoC modifications. All INC versions use a single parameter server (PS), while the software version uses eight to provide enough throughput.

Figure~\ref{fig:byteps} shows the average training speed per worker. We have the following observations:
1) INC solutions outperform non-INC ones for most models because they avoid incast to the PS. \sysname, ATP, and SwitchML are $42\%$, $42\%$, and $11\%$ faster than BytePS in VGG16;
2) For all models, \sysname performs similar to ATP ($97\%$ to $100\%$ of ATP), and at most $28\%$ faster than SwitchML; 3) the training speeds on ResNet are similar because they are computation-intensive, and communication does not affect the overall performance much. 

We believe the performance gain in \sysname over existing systems is from the automatic parallel streams. As a side benefit, \sysname uses only a single port (or one pipeline) instead of recirculation like ATP or SwitchML.  Using fewer ports is essential for the multi-application data plane. SwitchML-RDMA~\cite{switchmlRDMA} uses even more pipelines by chaining four pipelines together to achieve a performance gain over ATP. We do not adopt the design because resource efficiency is one of our key considerations.

\para{Paxos.} 
We use \sysname to implement a Paxos~\cite{lamport2001paxos} consensus system, offloading the leader and vote counting functions to switches.  The implementation only contains about 700 LoC changes.  We use an INC baseline, P4xos~\cite{dang2020p4xos},  and two software ones, \texttt{libpaxos}~\cite{libpaxos} and DPDK Paxos~\cite{dang2020p4xos}. 
We run two proposers, two acceptors, and three learners in all cases. 

Figure~\ref{fig:paxos} summarizes the results on both throughput and 99th-percentile latency to achieve one consensus. Key findings include:
1) \sysname achieves a maximum throughput of 503K messages/second, $12\%$ higher than P4xos, and $7.86\times$ and $4.93\times$ higher than the two software solutions. INC solutions are much faster because they offload packet processing to the switch to alleviate the CPU bottleneck on servers.
\sysname has higher throughput than P4xos because it only sends the final results to the learners, reducing the workload on servers and saving the traffic on learner links.
2) The 99th-percentile latency of \sysname is 311 ms and 96 ms shorter than software but 42 ms higher than P4xos.  This is because we choose not to run the acceptors on switches like P4xos and thus need an extra round trip to the software acceptor. We believe the location and replication flexibility of the acceptor is a worthwhile tradeoff for the extra latency, given that it is still much faster than pure software.

\begin{figure*}[tb]
	\begin{minipage}[t]{0.32\textwidth}
		\centering
		\includegraphics[width=1.0\columnwidth]{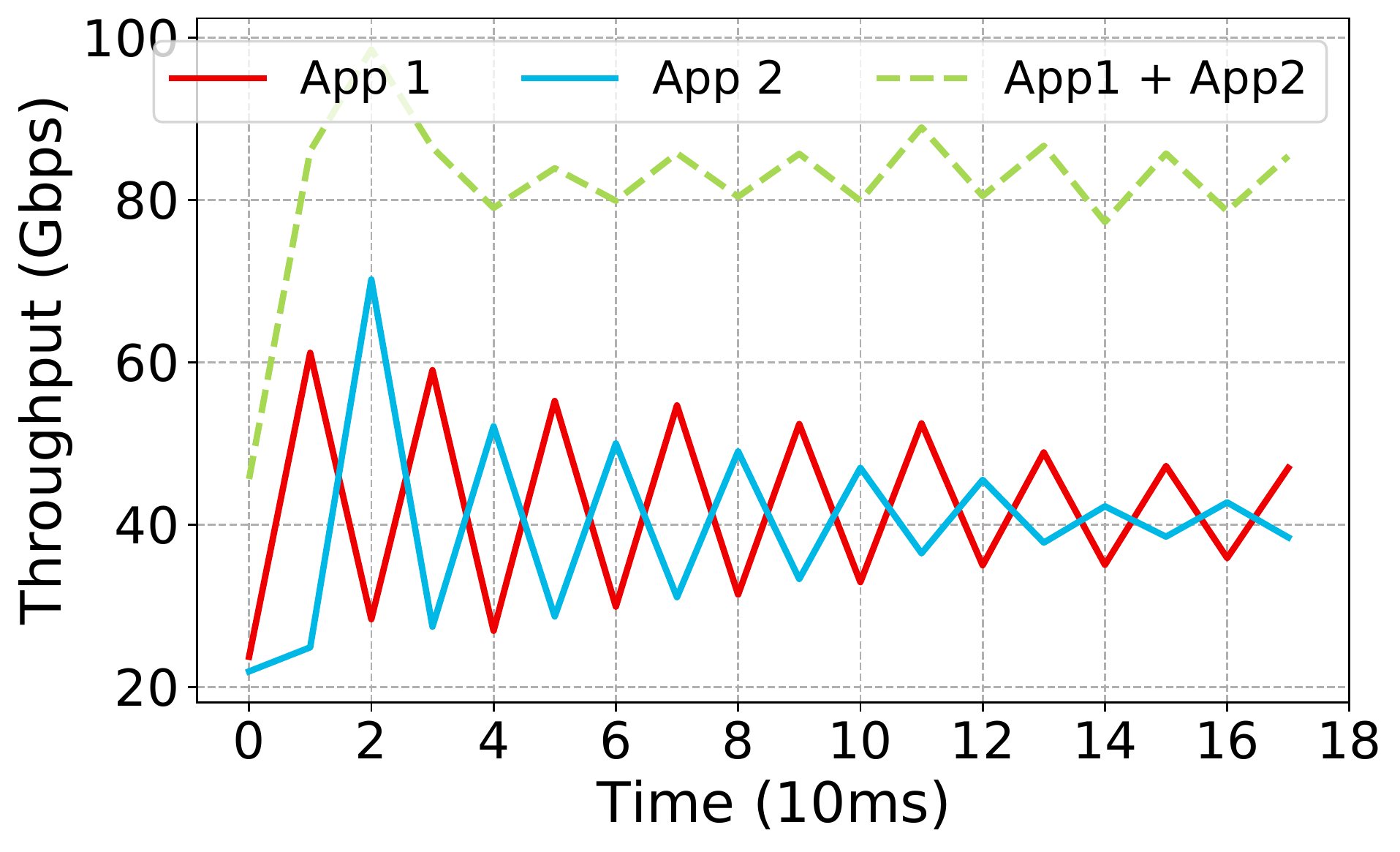}
		\caption{Congestion Control: Fairness}
		\label{fig:cc}
	\end{minipage}
	\hfill
	\begin{minipage}[t]{0.32\textwidth}
		\centering
		\includegraphics[width=1.0\columnwidth]{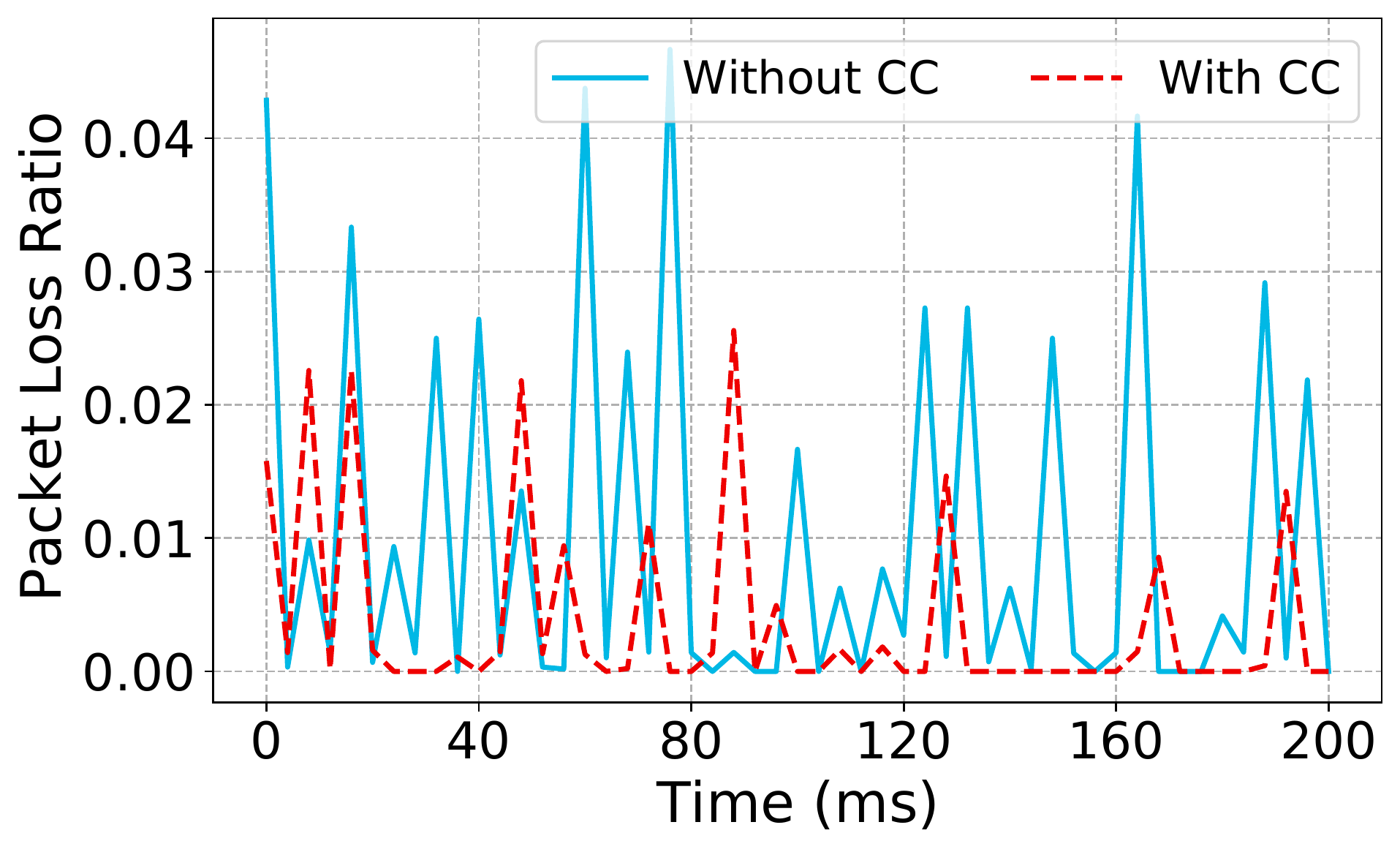}
		\caption{\small{Congestion Control: Packet Loss} }
		\label{fig:ccpl}
	\end{minipage}
	\hfill
	\begin{minipage}[t]{0.32\textwidth}
		\centering
		\includegraphics[width=1.0\columnwidth]{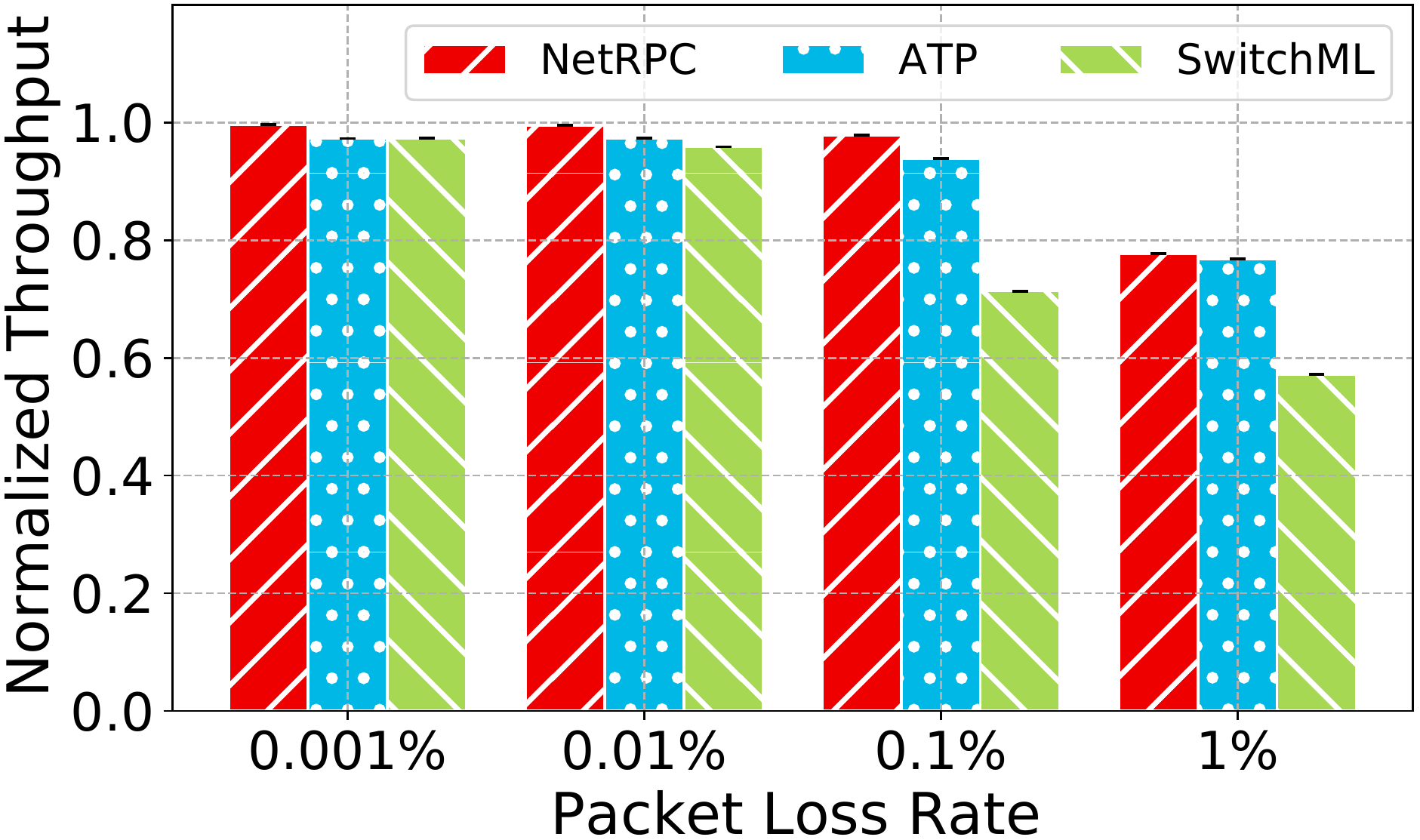}
		\caption{\small{Packet Loss Rates vs. Throughput}}
		\label{fig:plr}
	\end{minipage}
\end{figure*}

\begin{figure*}[tb]
	\begin{minipage}[t]{0.32\textwidth}
		\centering
		\includegraphics[width=1.0\columnwidth]{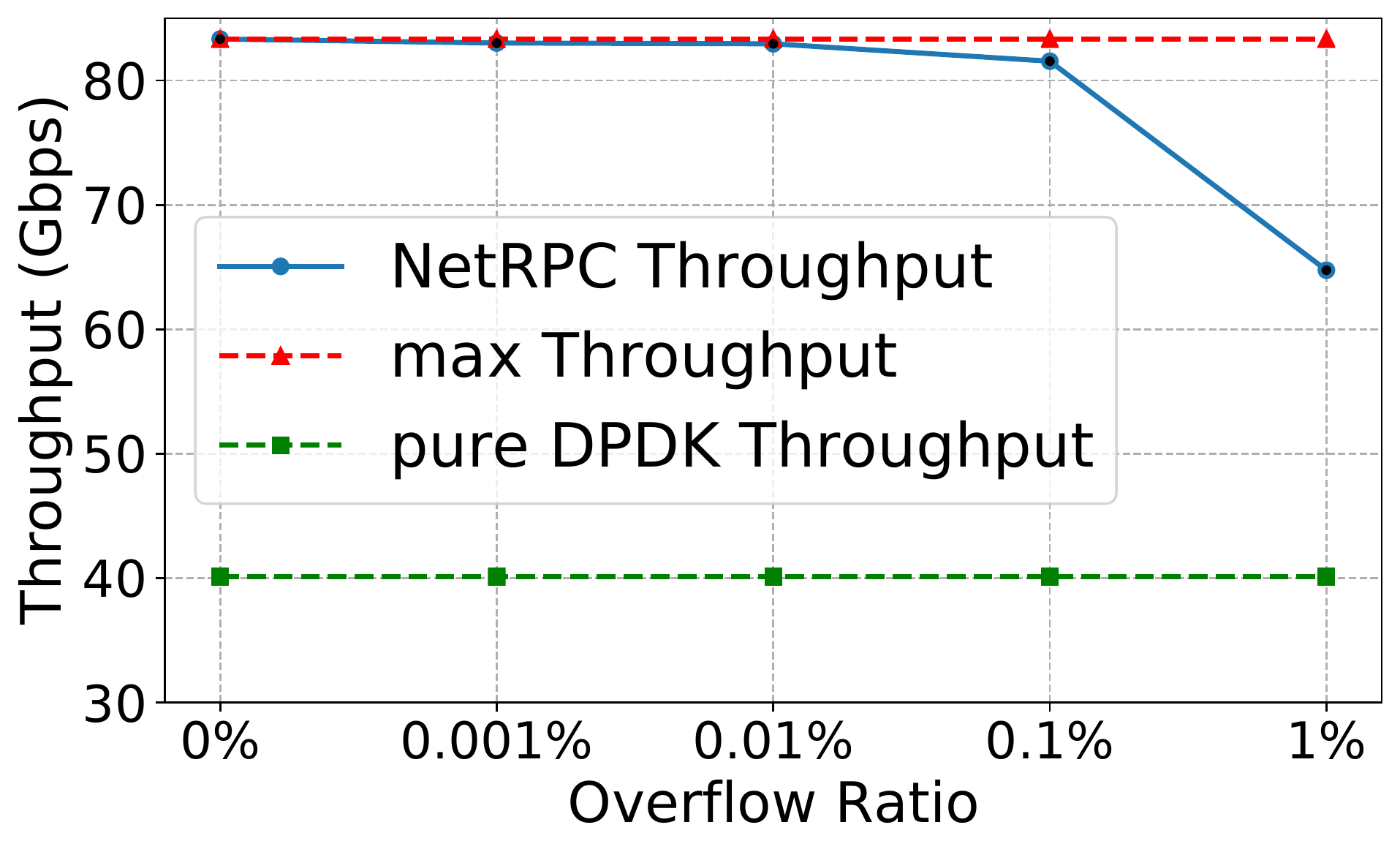}
		\caption{\small{Overflow Ratio vs. Throughput}}
		\label{fig:of}
	\end{minipage}
	\hfill
	\begin{minipage}[t]{0.32\textwidth}
		\centering
		\includegraphics[width=1.0\columnwidth]{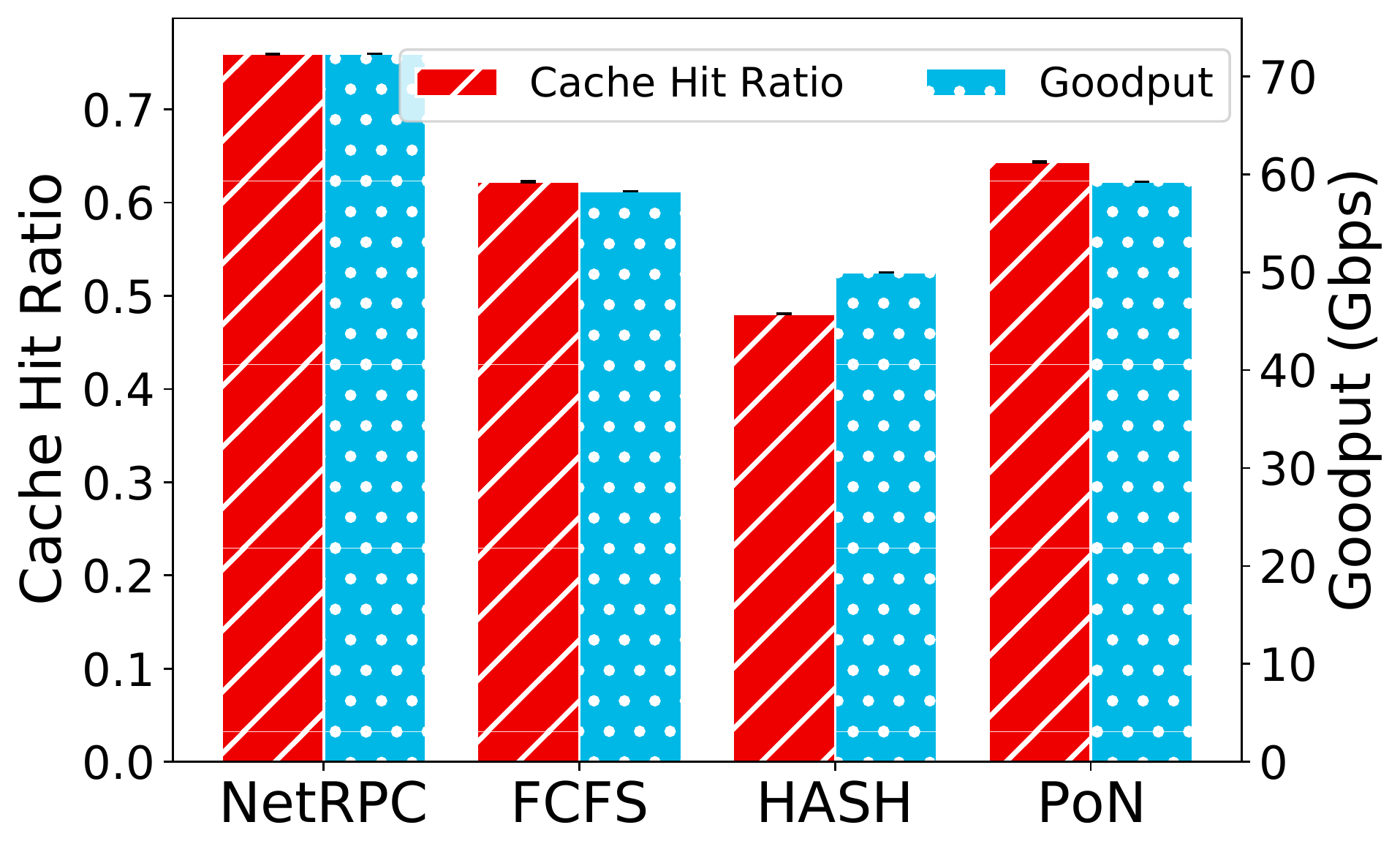}
		\caption{Caching Policy Comparison}
		\label{fig:cache}
	\end{minipage}
	\hfill
	\begin{minipage}[t]{0.32\textwidth}
		\centering
		\includegraphics[width=1.0\columnwidth]{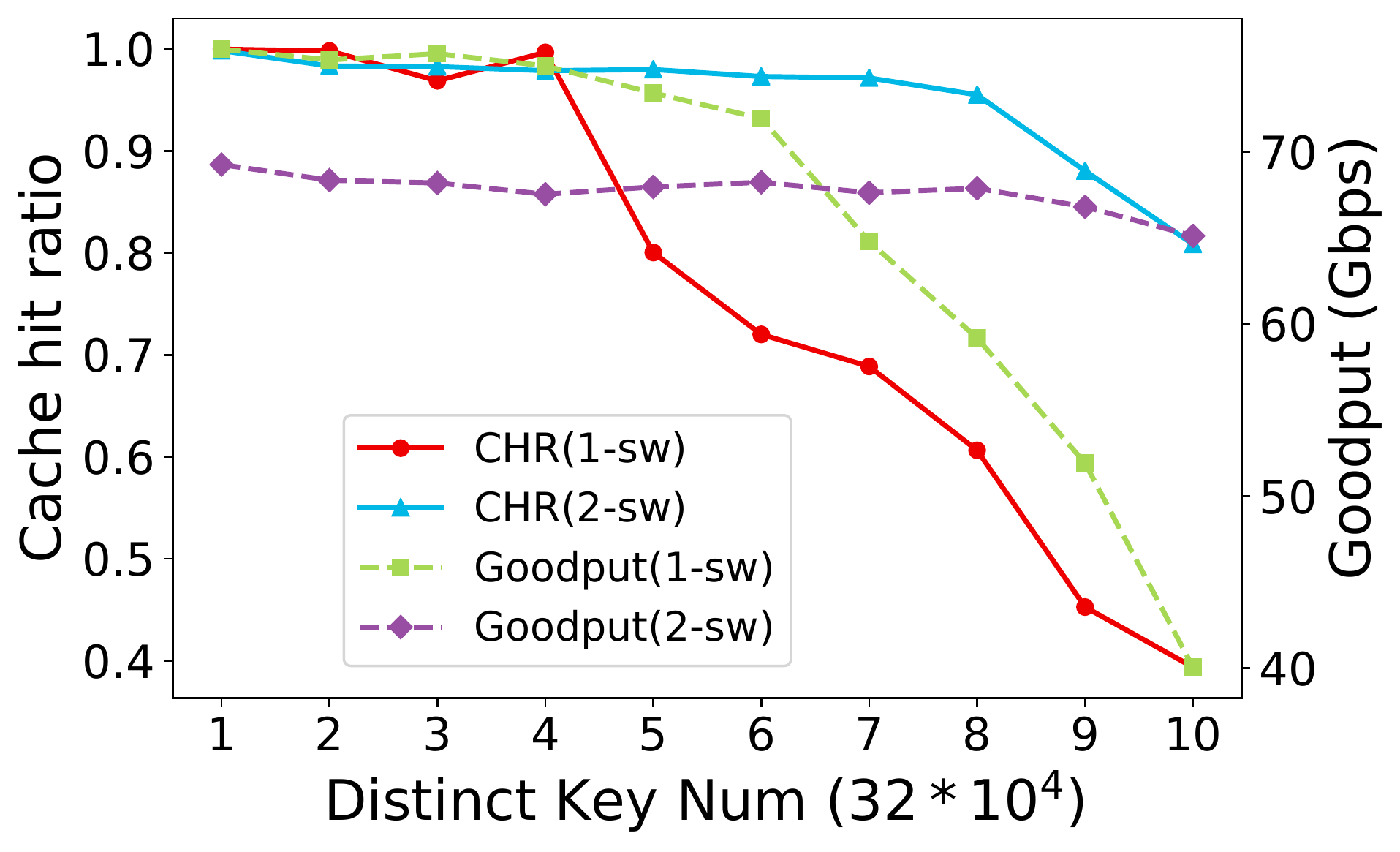}
		\caption{\sysname on Two Switches}
		\label{fig:pipe}
	\end{minipage}
\end{figure*}

\subsection{Micro-benchmarks}

To better understand \sysname performance impact, we conduct a series of micro-benchmarks, focusing on INC-related functions only. We also use both prior INC arts and pure software DPDK implementation as comparison baselines.

\begin{table}[tb]
	\caption{Microbenchmark on Basic INC Functions}
	\label{tab:microbenchmark}
		
\centering
\footnotesize
        \begin{tabular}{|c|c|c|c|}
			\hline
			Metrics             & \sysname & Prior Arts & DPDK  \\ \hline
			SyncAgtr Goodput(Gbps)  & 50.55  & 46.44 (ATP)     & 40.11 \\ \hline
			AsyncAgtr Goodput(Gbps) & 72.31  & 73.96 (ASK)    & 45.88 \\ \hline
			Voting Delay($\mu$s)     & 20     & 22 (P4xos)       & 92    \\ \hline
			Monitor Delay(ms)   & 3.52   & 3.26 (ElasticSketch)  & 4.05  \\ \hline
			\begin{tabular}[c]{@{}c@{}}Packet Processing \\ Capacity(Mpps)\end{tabular} & \textgreater{}1000 & \textgreater{}1000 & 83.47 \\ \hline
		\end{tabular}%
\end{table}

\para{Throughput.} 
We perform SyncAtgr and AsyncAtgr on a 2-to-1 testbed and measure the \emph{sender goodput}, using ATP and ASK as INC baselines. 

The first row in Table~\ref{tab:microbenchmark} shows the result.  \sysname offers $9\%$ higher throughput than ATP. The reason is that \sysname does not apply recirculation (we use \code{copy} policy in this experiment) as ATP and SwitchML, which costs extra ports or pipelines on the switch . Instead, it relies on the parallel message sending (Section~\ref{sec:parallel}) to increase the goodput. Not surprisingly, both INC solutions outperform software solutions, e.g., \sysname offers $26\%$ higher goodput than pure DPDK. In fact, the end-to-end training results ($42\%$ faster, see Section~\ref{sec:macrobenchmark}) show an even larger improvement than the micro-benchmark, as in SyncAtgr, the shorter latency also improves GPU utilization as we spend less time waiting for the aggregation results. 

The second row shows the goodput in AsyncAtgr. \sysname achieves a similarly high throughput as ASK (about 73 Gbps). Unlike SyncAtgr, the keys count as part of a valid payload in this case, and thus the goodput is higher.  
Both INC solutions have $37\%$ higher throughput than the pure DPDK.

\para{Latency.  } 
We measure the average latency for the two latency-sensitive applications: Agreement and KeyValue, using P4xos voting and ElasticSketch~\cite{yang2018elastic} (monitoring) as baselines.
The third row in Table~\ref{tab:microbenchmark} shows the average voting latency. Both \sysname and P4xos outperform DPDK with a $76\%$ latency reduction.  \sysname and P4xos offer similar latency, showing that \sysname abstraction layers do not add extra latency. 

The last two rows of Table~\ref{tab:microbenchmark} compares performance for KeyValue types, specifically in \emph{flow counting}.  
Both \sysname and ElasticSketch have lower latency than DPDK, by $13\%$ and $20\%$, respectively. Notably, the last row of Table~\ref{tab:microbenchmark} also shows a $10\times$ packet processing capacity increase from DPDK.  
\sysname is about 0.26 ms ($9\%$) slower than ElasticSketch because we do not have the same application-specific optimization that avoids modifying packets. We believe a less-than-$10\%$ latency increase is a reasonable price to pay for the general programming model by omitting the optimization.

\para{Congestion control performance.}  
To evaluate the effects of congestion control in \sysname, 
we concurrently run two applications: a SyncAggr and an AsyncAggr on the same data plane (i.e., the same switch, host, and links), each having two clients and one server. 
Figure~\ref{fig:cc} shows the throughput over a short time period. We observe that the throughput quickly converges within 200 ms, and the combined bandwidth reaches $77\%$ to $89\%$ of the 100Gbps link. Also, the two application fairly shares the available bandwidth.  
Figure~\ref{fig:ccpl} shows the packet loss ratio over a short time period with/without congestion control. We can see that our ECN-based congestion and flow control reduces packet loss by about $63\%$, as it automatically adjusts the sending window to avoid overwhelming both the link and the server agent (Section~\ref{sec:cc}).

\para{Reliability mechanisms. } 
To evaluate how \sysname handles packet losses, 
we inject packet losses at different rates to emulate unreliable network. 
We run three INC applications \sysname, ATP, and SwitchML and verified that all three correctly handles packet loss. Figure~\ref{fig:plr} shows the normalized throughput. 
\sysname performs retransmission correctly under packet loss, using on-switch states only. At a high loss rate, \sysname has a more graceful performance degradation. 
Compared with the no-loss case, \sysname, ATP, and SwitchML's throughput decrease by $22\%$, $23\%$, and $43\%$, respectively. With $1\%$ loss, \sysname shows significantly less performance degradation than SwitchML because it adopts out-of-order ACKs and thus learns and reacts to packet loss faster.

\para{Handling overflows.  } 
We run SyncAggr under synthetic workload varying overflow ratios from $0.001\%$ to $1\%$. Figure~\ref{fig:of} plots the throughput vs. overflow ratios. 
In all experiments, we check the computation results to ensure that \sysname detects and corrects the overflow as we expect.
When the overflow ratio exceeds $0.1\%$, we notice throughput degradation due to the software fallback. 
\sysname still achieves about $65$ Gbps throughput at $1\%$ overflows. Note that the overflow ratio in real workload is far less than  
$1\%$ with a reasonable quantization scaling factor for floating-point numbers. 
In contrast, the pure software solution only achieves a max of 40 Gbps.

\begin{table}[tb]
\caption{Clear Policy Impact on Performance}
\label{tab:clear}
\centering
\begin{tabular}{|c|c|c|c|}
\hline
                  & Latency & Memory & Throughput \\ \hline
copy    & 74$\mu$s    & 1x     & 83.11Gbps  \\ \hline
shadow       & 24$\mu$s    & 2x     & 50.41Gbps  \\ \hline
lazy (0\%)  & 22$\mu$s    & 1x     & 83.31Gbps  \\ \hline
lazy (1\%)  & 23$\mu$s    & 1x     & 64.75Gbps  \\ \hline
lazy (10\%) & 30$\mu$s    & 1x     & 34.82Gbps  \\ \hline
\end{tabular}%
\end{table}

\para{Performance of \code{clear} policies.} 
\sysname offers three ways to handle \pclear in \netfilter (Section~\ref{sec:memory}). 
We measure the performance of a 2-to-1 SyncAggr using three \pclear policies, and Table~\ref{tab:clear} summarizes the results.  
\code{Lazy} policy performance depends on the ratio of arithmetic overflow, and we use three ratios of $0\%$, $1\%$, and $10\%$.
\code{Copy} policy achieves the highest throughput without extra memory cost but also has the highest latency because it relies on servers to backup the cleared states for reliability. 
\code{Shadow} policy offers a good latency of $24 \mu s$ but doubles memory usage and has the lowest throughput because it needs to recirculate the packet and keep an extra copy. 
\code{Lazy}  policy achieves both the highest throughput and lowest latency with no overflows. But as the overflow ratio increases, both metrics degrade. The actual accumulator overflow ratio depends on the data. Thus, we leave them as a user configuration in the \netfilter. 
 
\para{Cache policy.} As we discuss in Section~\ref{sec:memory}, a good cache policy alleviates traffic incast at the server and improves performance. We evaluate multiple cache policies. The experiment uses $32 \times 4K$ switch memory with 2-to-1 traffic. Comparison baselines are FCFS, hash-based caching (HASH), and Power of $N$ (PoN). HASH policy uses the hash key as the index to address the switch memory (like ASK \cite{ASK} and ATP \cite{265053ATP}) and falls back to the server agent on hash collisions. PoN is a classic policy in sketches \cite{yang2018elastic}: it only caches the hot keys whose hit number exceeds a threshold $N$ and gives up caching when the switch memory is full. We tune the hyperparameter $N$ to maximize the performance experimentally. 

Figure~\ref{fig:cache} shows the result. 
First, the CHR is positively correlated with the goodput, indicating the need for cache policy optimization. 
\sysname's periodic cache update outperforms other cache policies by $18\% \sim 57\%$ on cache hit ratio (CHR) and $22\% \sim 44\%$ on goodput. 
HASH performs the worst because it ignores the locality of keys in the same packet: if some keys are cached, but their adjacent keys in the same stream are not due to hash collision, the entire packet will never hit the cache. 
PoN and FCFS behave similarly as they stop caching new hot keys if the cache has been fully filled.
Compared with these baselines, \sysname catches up the locality better and adapts to high-skewed key distribution better because it always caches the recent hot keys and periodically updates the switch cache to make up space for newer ones.

\begin{table}[tb]\footnotesize
\caption{Concurrent Application Throughput and Latency}
\label{tab:concurrent1}
\resizebox{\columnwidth}{!}{%
\begin{tabular}{|c|c|c|c|c|}
\hline
Metrics             & 1APP  & 4APP  & 4APP$\times5$ \\ \hline
Sync Goodput (Gbps)  & 50.55 & 24.88 & 24.84      \\ \hline
Async Goodput (Gbps) & 72.31 & 36.01 & 36.60     \\ \hline
Goodput Sum (Gbps)   & N/A      & 60.89 & 61.44      \\ \hline \hline
KeyValue Delay (ms) & 3.52 & 3.56 & 3.85       \\ \hline 
Agreement Delay ($\mu$s) & 20   & 21   & 24          \\ \hline
\end{tabular}%
}
\end{table}

\subsection{Multiple Concurrent Applications}

An important goal of  \sysname is to support a multi-application data plane without switch rebooting. To evaluate the performance,
we run multiple instances of all four application types in a 2-to-1 topology.  
We evaluate using three concurrency settings: 1) running a single application instance (``1APP ''); 2) running one instance per type (``4APP''); and 3) running five instances per type (``4APP $\times 5$'') . 
Table~\ref{tab:concurrent1} shows the total goodput and average latency.
In all cases, we measure and report the throughput of SyncAgtr and AsyncAgtr and the latency of KeyValue and Agreement. In the 4APP $\times 5$ case, we take the average of all instances of the measured type. 

When concurrent applications increase from 4 to 20, we observe that the total bandwidth of SyncAgtr and AsyncAgtr stays roughly the same.  
Although KeyValue and Agreement do not use much bandwidth, they do contend for switch PPS (packets per second) and queue up in sending threads. The experiments show that small applications have little impact on bandwidth-heavy ones. We observe only a $20\%$ latency increase compared to the 1APP case. These results demonstrate the successful resource sharing ability of \sysname.

\subsection{Running on Multiple Switches}

Limited by available hardware, we only validate \sysname's cross-switch capability with two-switches. We chain the two switches into a longer pipeline, and thus a packet can carry more key-value pairs. The \sysname server agent decides which key to put on which switch. We compare the performance of running 2-to-1 MapReduce on the testbed with one / two switches. We loop through the distinct keys multiple times, and thus a cache smaller than the number of distinct keys will suffer cache misses. Then we measure the CHR and the goodput varying with the number of distinct keys as an indicator of how well \sysname is using memory on both switches. 

Figure~\ref{fig:pipe} shows the result. Each switch stores $M = 32 \times 40K$ values with distinct keys.  We confirm that the goodput starts to drop at $M$ using one switch, but $2M$ with two. The peak goodput decreases slightly with more switches (from $75$ Gbps to $69$ Gbps), mainly because of the increased host workload to encode more keys into the packet. 
Beyond the switch memory capacity, the goodput first decreases slightly ($5.3\%$ of peak throughput with $1.5M$ keys for one switch) and then dramatically ($22\%$ with $2M$ keys). 
This is because offering a $75$~Gbps workload, there is little hope that 
the server CPU can handle many cache misses.
Nevertheless, the two-switch setting shows a $1.63\times$ improvement over the one-switch case when handling $2.5M$ distinct keys, showing that \sysname can efficiently utilize memory on multiple switches.

\section{Conclusion and Future Work}
\label{sec:conclusion}

In-network computation (INC) comes from software-defined networking (SDN), but INC is fundamentally different from SDN because it mainly provides \emph{computation} service instead of \emph{communication}. Thus, we need a new programming model for INC to \emph{better describe computation}. We need high-level data structures, collections, memory, and procedure calls that center around end-hosts instead of packets, headers, tables, and pipelines that center around switches. On the other hand, we recognize that the INC data plane is still a shared network infrastructure, not an application-specific accelerator. Thus, both generality and multi-application support are essential.  

\sysname, to our knowledge, is the first framework that integrates INC into the familiar RPC programming model.  \sysname allows users to implement different types of INC applications using the familiar gRPC framework and run them on a single shared INC data plane.  \sysname achieves 97\% of LoC deduction for INC applications and offers similar or better performance boosts than handcrafted systems.

Current \sysname mainly focuses on \emph{mechanisms} of INC + RPC integration.  In future work, we will focus on \emph{policies}, such as scheduling among different applications, efficient sharing between INC workload and other SDN or traditional network traffic, efficient end-host CPU, GPU, and INC co-scheduling. We will also explore \sysname on more complex topologies, especially those with oversubscribed links. 
We will extend \sysname congestion control with more fine-grained window adjustment.
We will open source \sysname on the publication of this paper to benefit the INC community.




\bibliographystyle{plain}
\bibliography{reference}

\appendix
\section{Arithmetic in \titlesysname}
\label{sec:op}
We list the arithmetic operators of \pmodify and their semantics supported by \sysname in Table~\ref{tab:op}.
\begin{table}[tb]
\caption{Arithmetic Operations in \pmodify}
\footnotesize
\label{tab:op}
\resizebox{0.9\columnwidth}{!}{%
\begin{tabular}{|c|c|}
\hline
\textbf{OP}     & \textbf{Semantics}                                       \\ \hline
MAX    & stream.value = max(stream.value, para)          \\ \hline
MIN    & stream.value = min(stream.value, para)          \\ \hline
ADD    & stream.value += para                            \\ \hline
ASSIGN & stream.value = para                             \\ \hline
SHIFTL & stream.value \textless{}\textless{}= para       \\ \hline
SHIFTR & stream.value \textgreater{}\textgreater{}= para \\ \hline
BAND   & stream.value \&= para                           \\ \hline
BOR    & stream.value |= para                            \\ \hline
BNOT   & stream.value = $\sim$stream.value               \\ \hline
BXOR   & stream.value \textasciicircum{}= para           \\ \hline
\end{tabular}%
}
\end{table}

\section{\titlesysname Protocol}

\begin{figure}[tb]
	\centering
	\includegraphics[width=1.0\columnwidth]{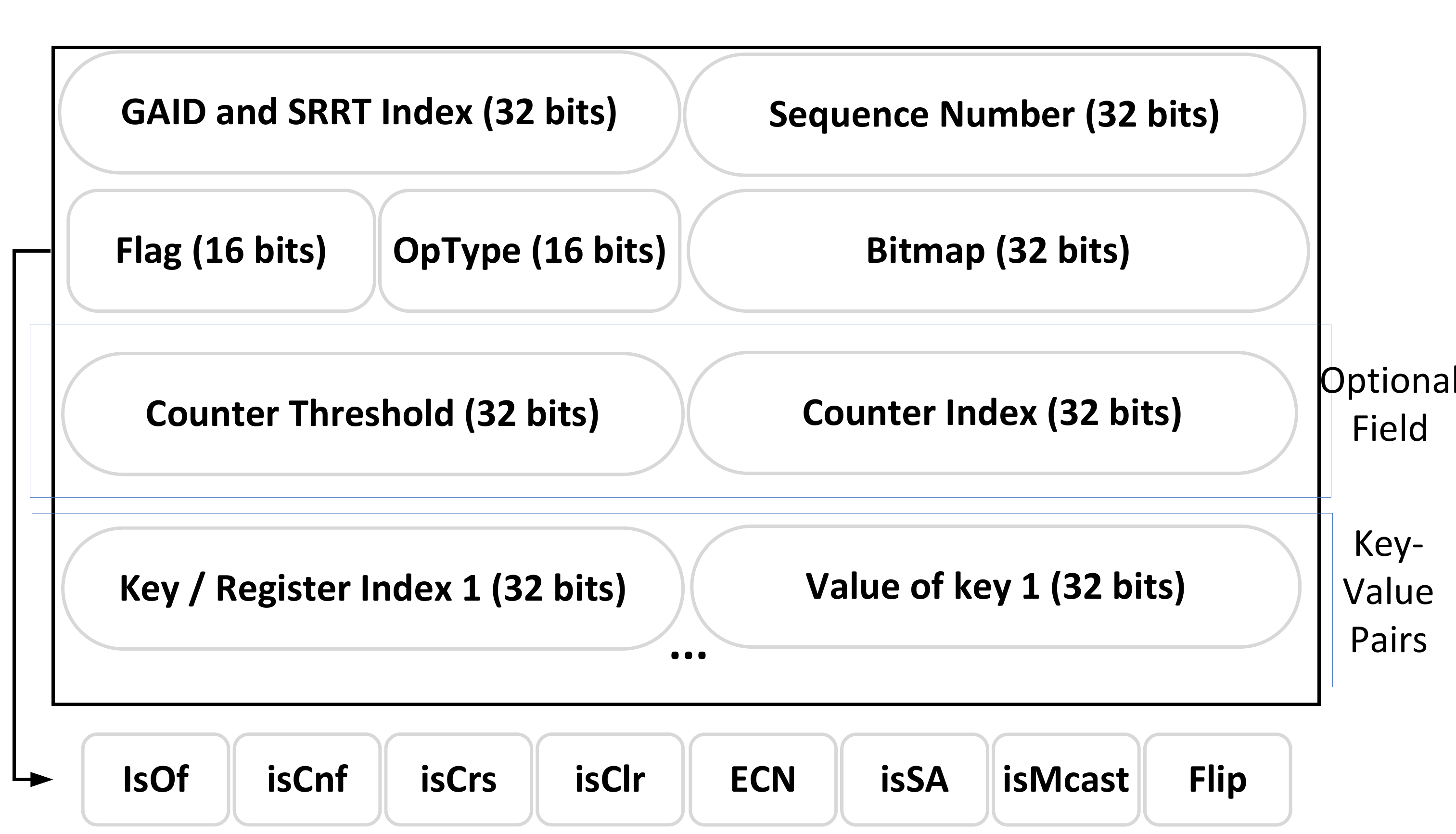}
	\caption{\sysname Packet Format}
	\label{fig:header}
\end{figure}

\subsection{Packet Format} 
\label{sec:header}
The packet contains three kinds of fields. The key-value pairs encode the application data with the format of an array of <key/index, value> tuples; computation control fields encode the \netfilter configurations and guide the switch program for the computation; transport control fields maintain the channel connection.

\para{Key-value pairs.}
Each \sysname packet carries 32 key-value pairs. These pairs are either processed on the switch or the server agent by the selected primitives. The computation results are also carried back in the same position. 

\para{Computation Control Fields.} The control flag bits contain the basic information about primitives selection. Current bits in use can indicate the following choices: whether any overflow happens \texttt{(isOf}); whether to use \pcntforward (\texttt{isCnf}); whether to clear the target memory (\texttt{isClr}).

\texttt{OpType} indicates the type of arithmetic operation on key-value pairs. \sysname supports various line-rate on-packet computation as we discuss in Appendix~\ref{sec:op}. 
In \texttt{bitmap} field, the $i$-th bit in the bitmap indicates whether the switch should process the $i$-th key-value pair. 
The \pcntforward fields only come into effect when the \texttt{isCnf} flag is set. \texttt{counter index} tells the switch which counter (register) to increase; when the register value equals to the \texttt{counter threshold}, the switch should forward the packet instead of dropping it.

\para{Transmission Control Fields.} Concurrent \sysname connections (de)multiplex the network, and  \sysname distinguishes the flows by the {\tt GAID}. On hosts, received packets are classified to the applications; on the switch, the {\tt GAID} is also used for admission control. In \sysname, each sending thread maintains a short-term connection to serve applications' calls/tasks and thus assigns a sequence number (starting from zero) for each packet. In addition, the reliability control requires sending threads to maintain a long-term connection (cross the tasks) with the switch. The field State Register of Reliable Transmission {\tt SRRT} is the switch memory address to store the state, and the {\tt flip bit} is the reliable state to store.
Some bits in the \texttt{Control Flag} also controls the routing: whether the packet should cross the switch to the server agent (\texttt{isCross}); \texttt{ECN} indicates whether the switch is experiencing congestion (queue buildup); whether the packet comes from the server agent (\texttt{isSA}); whether to multicast the packet (\texttt{isMcast}).

\para{Optimization.} Some optional fields will be removed if unnecessary in the computation to improve the network bandwidth efficiency and the goodput.
(1) If we address the key-value or value stream linearly to the switch memory, we can eliminate the key fields and indicate the starting index of the memory segment by the {\tt counter index} field. (2) If the computation does not need \pcntforward, we can eliminate the \pcntforward fields.

\section{Switch pipeline details }
\label{sec:switch-pipeline}

\begin{figure}[tb]
	\centering
	\includegraphics[width=1.0\columnwidth]{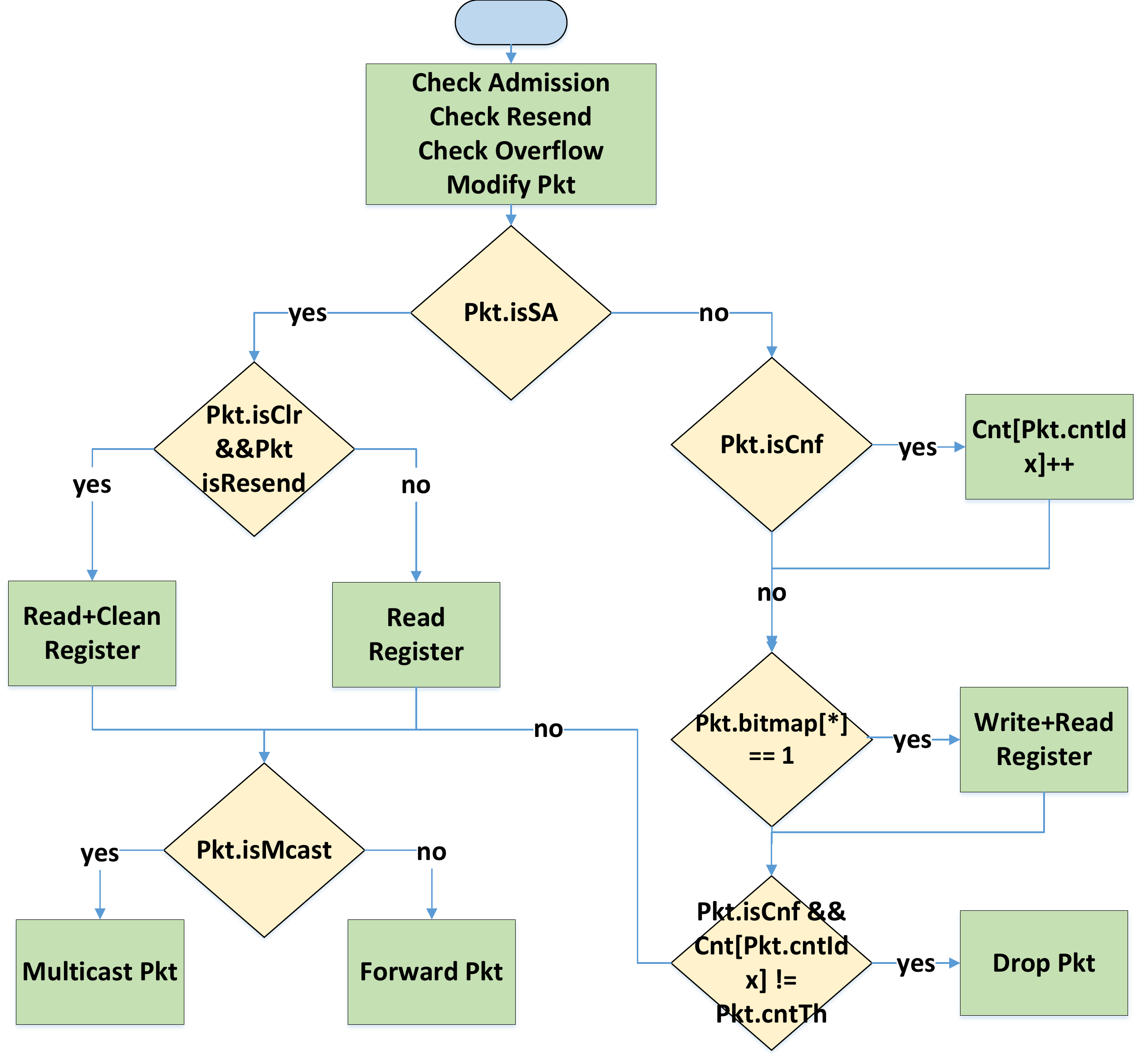}
	\caption{\sysname Switch Logic}
	\label{fig:logic}
\end{figure}

There is a 12-stage pipeline in our switch, and we use 8 to implement the map access primitives. The remaining four stages handle the reliable transmission, flow and congestion control, as well as \pmodify and the \pcntforward primitive. Figure~\ref{fig:logic} illustrate the flowchart for switch logic. 

When the switch receives a \sysname packet, it will first check whether the corresponding application (GAID) has registered. Unregistered packets will be forwarded as normal ones. Moreover, the switch checks whether it receives the packet for the first time. Otherwise, it avoids \pread/\pclear primitives on the switch memory but still \pwrite values from registers into the packets. An overflow packet will be forwarded directly for fallback without on-switch processing. 

For packets to the server, the switch first executes \pmodify and \pcntforward if required, then processes key-value pairs in the packet: \pread the switch registers and \pwrite the computation results back to replace the value. The switch drops those packets that enable \pcntforward but do not reach the threshold and forwarded/multicast the rest packets.

For packets from the server, the switch first \pwrite register values into the packet and then decides whether to clear the corresponding registers. The switch will forward/multicast the packets according to control flags and routing rules.

\section{\titlesysname Implementation Examples}
\label{appendix:example}
We enumerate some \sysname implementation of classic INC applications:
MapReduce, lock server, and network monitoring in Figure~\ref{code:define1} to \ref{code:example3.1}.

\begin{figure}[htb]\centering\setlength{\abovecaptionskip}{0.cm}\begin{minipage}[t]{0.85\columnwidth}
\begin{lstlisting}[language=C++]
import "netrpc.proto"
message ReduceRequest {
  netrpc.STRINTMap kvs = 1;
}
message ReduceReply {
  string msg = 1;
}
message QueryRequest {
  string msg = 1;
}
message QueryReply {
  netrpc.STRINTMap kvs = 1;
}
service MapReduce {
  rpc ReduceByKey (ReduceRequest) returns (ReduceReply) {} filter "reduce.nf"
  rpc Query (QueryRequest) returns (QueryReply) {} filter "query.nf"
}
\end{lstlisting}
\caption{RPC Service Definition of Distributed MapReduce}
\label{code:define1}
\end{minipage}\end{figure}

\begin{figure}[htb]\centering\begin{minipage}[t]{0.85\columnwidth}
\begin{lstlisting}
{ //reduce.conf
  "AppName": "MR-1",
  "Precision": 0,
  "get": "nop",
  "addTo": "ReduceRequest.kvs",
  "clear": "nop",
  "modify": "nop",
  "CntFwd": {
    "to": "SRC", 
    "threshold": 0, 
    "key": "NULL",
  },
}
{ //query.conf
  "AppName": "MR-1",
  "Precision": 0,
  "get": "QueryReply.kvs",
  "addTo": "nop",
  "clear": "nop",
  "modify": "nop",
  "CntFwd": {
    "to": "SRC", 
    "threshold": 0, 
    "key": "NULL",
  },
}
\end{lstlisting}
    \caption{\netfilter of Distributed MapReduce}
    \label{code:conf1}
\end{minipage}\end{figure}

\begin{figure}[htb]\centering\begin{minipage}[t]{0.85\columnwidth}
\begin{lstlisting}[language=C++]
shared_ptr<Channel> channel = CreateCustomChannel(server_ip, InsecureChannelCredentials());
unique_ptr<Stub> stub_(NewStub(channel));
pair<string,int>* MapReduce(pair<string,int>* data, int length) {
    ReduceRequest request1;
    ReduceReply reply1;
    ClientContext context1;
    for(int i = 0; i<length; i++){
        (*request1.mutable_kvs()->mutable_map())[data[i].first] = data[i].second;
    }
    Status status = stub_->ReduceByKey(&context1, request1, &reply1);
    QueryRequest request2;
    QueryReply reply2;
    ClientContext context2;
    stub_->Query(&context2, request2, &reply2);
    int sz = reply2.mutable_kvs()->mutable_map()->size(), idx = 0;
    pair<string,int>* output = new pair<string,int>[sz];
    for(auto it: (*reply2.mutable_kvs()->mutable_map())){
        output[idx].first = it.first;
        output[idx++].second = it.second;
    }
    return output;
}
\end{lstlisting}
\caption{Client Stub for Distributed MapReduce}
\label{code:example1}
\end{minipage}\end{figure}

\begin{figure}[htb]\centering\begin{minipage}[t]{0.85\columnwidth}
\begin{lstlisting}[language=C++]
import "netrpc.proto"
message LockRequest {
  netrpc.STRINTMap map = 1;
}
message LockReply {
  string msg = 1;
}
message ReleaseRequest {
  netrpc.STRINTMap map = 1;
}
message ReleaseReply {
  string msg = 1;
}
service Lock {
  rpc GetLock (LockRequest) returns (LockReply) {} filter "lock.nf"
  rpc Release (ReleaseRequest) returns (ReleaseReply) {} filter "release.nf"
}
\end{lstlisting}
\caption{RPC Service Definition of Distributed Lock Server}
\label{code:define2}
\end{minipage}\end{figure}

\begin{figure}[htb]\centering\begin{minipage}[t]{0.85\columnwidth}
\begin{lstlisting}
{ //lock.conf
  "AppName": "LS-1",
  "Precision": 0,
  "get": "nop",
  "addTo": "nop",
  "clear": "nop",
  "modify": "nop",
  "CntFwd": {
    "to": "SRC", 
    "threshold": 1, 
    "key": "LockRequest.kvs",
  },
}
{ //release.conf
  "AppName": "LS-1",
  "Precision": 0,
  "get": "nop",
  "addTo": "nop",
  "clear": "copy",
  "modify": "nop",
  "CntFwd": {
    "to": "SRC", 
    "threshold": 0, 
    "key": "ReleaseRequest.kvs",
  },
}
\end{lstlisting}
    \caption{\netfilter of Distributed Lock Server}
    \label{code:conf2}
\end{minipage}\end{figure}

\begin{figure}[htb]\centering\begin{minipage}[t]{0.85\columnwidth}
\begin{lstlisting}[language=C++]
shared_ptr<Channel> channel = CreateCustomChannel(server_ip, InsecureChannelCredentials());
unique_ptr<Stub> stub_(NewStub(channel));
void BlockingLock(string* lockTarget, int length) {
    LockRequest request1;
    LockReply reply1;
    ClientContext context1;
    for(int i = 0; i<length; i++){
        (*request1.mutable_kvs()->mutable_map())[lockTarget[i]] = 1;
    }
    Status status = stub_->LockSend(&context1, request1, &reply1);
    /* critical section */
    ReleaseRequest request2;
    ReleaseReply reply2;
    ClientContext context2;
    for(int i = 0; i<length; i++){
        (*request2.mutable_kvs()->mutable_map())[lockTarget[i]] = 0;
    }
    stub_->Release(&context2, request2, &reply2);
}
\end{lstlisting}
\caption{Client Stub for Blocking Lock Acquire and Release}
\label{code:example2}
\end{minipage}\end{figure}

\begin{figure}[htb]\centering\begin{minipage}[t]{0.85\columnwidth}
\begin{lstlisting}[language=C++]
import "netrpc.proto"
message MonitorRequest {
  netrpc.STRINTMap kvs = 1;
  string payload = 1;
}
message MonitorReply {
  string payload = 1;
}
message QueryRequest {
  string message = 1;
}
message QueryReply {
  netrpc.STRINTMap kvs = 1;
}
service Monitor {
  rpc MonitorCall (MonitorRequest) returns (MonitorReply) {} filter "monitor.nf"
  rpc Query (QueryRequest) returns (QueryReply) {} filter "query.nf"
}
\end{lstlisting}
\caption{RPC Service Definition of Network Monitoring}
\label{code:define3}
\end{minipage}\end{figure}

\begin{figure}[htb]\centering\setlength{\abovecaptionskip}{0.cm}\begin{minipage}[t]{0.85\columnwidth}
\begin{lstlisting}
{ //monitor.conf
  "AppName": "MON-1",
  "Precision": 0,
  "get": "nop",
  "addTo": "MonitorRequest.kvs",
  "clear": "nop",
  "modify": "nop",
  "CntFwd": {
    "to": "SERVER", 
    "threshold": 0, 
    "key": "NULL",
  },
}
{ //query.conf
  "AppName": "MON-1",
  "Precision": 0,
  "get": "QueryReply.kvs",
  "addTo": "nop",
  "clear": "nop",
  "modify": "nop",
  "CntFwd": {
    "to": "SRC", 
    "threshold": 0, 
    "key": "NULL",
  },
}
\end{lstlisting}
    \caption{\netfilter of Network Monitoring}
    \label{code:conf3}
\end{minipage}\end{figure}

\begin{figure}[htb]\centering\setlength{\abovecaptionskip}{0.cm}\begin{minipage}[t]{0.85\columnwidth}
\begin{lstlisting}[language=C++]
shared_ptr<Channel> channel = CreateCustomChannel(server_ip, InsecureChannelCredentials());
unique_ptr<Stub>stub_(NewStub(channel));
pair<string,int>* MonitorRPC(string* metrics, int length) {
    MonitorRequest request1;
    MonitorReply reply1;
    ClientContext context1;
    for(int i = 0; i<length; i++){
        (*request1.mutable_kvs()->mutable_map())[metrics[i].first] = 1;
    }
    request1.payload = "Hello";
    Status status = stub_->MonitorCall(&context1, request1, &reply1);
    if (status.ok()) {
      cout << reply1.payload << endl;
    } 
    QueryRequest request2;
    QueryReply reply2;
    ClientContext context2;
    stub_->Query(&context2, request2, &reply2);
    int sz = reply2.mutable_kvs()->mutable_map()->size(), idx = 0;
    pair<string,int>* output = new pair<string,int>[sz];
    for(auto it: (*reply2.mutable_kvs()->mutable_map())){
        output[idx].first = it.first;
        output[idx++].second = it.second;
    }
    return output;
}
\end{lstlisting}
\caption{Client Stub for RPC with Monitoring}
\label{code:example3.1}
\end{minipage}\end{figure}


\end{document}